%% file: BG_CH.tex
\documentclass[a4paper,12pt]{article}
\usepackage{jheppub,booktabs}
\usepackage{amsmath,amssymb}
\usepackage{xcolor,colortbl}
\usepackage{cancel}
\usepackage{subfigure}
\usepackage{mathrsfs}
\usepackage{graphicx}
\usepackage{multicol}
\usepackage{multirow}
\usepackage{pdflscape}
\usepackage{hepunits}
\usepackage{feynmf}
\usepackage{enumerate}
\usepackage[utf8x]{inputenc}

\author{Andrei Angelescu,}
\author{Florian Goertz,}
\author{and Aika Tada}
\affiliation{Max-Planck-Institut f{\"u}r Kernphysik\\ Saupfercheckweg 1, 69117 Heidelberg, Germany}
\emailAdd{andrei.angelescu@mpi-hd.mpg.de}
\emailAdd{florian.goertz@mpi-hd.mpg.de}
\emailAdd{aika.tada@mpi-hd.mpg.de}

\title{\boldmath{$Z_2$} Non-Restoration and Composite Higgs: {\Large Singlet-Assisted Baryogenesis w\!/\!o \!Topological Defects}}

\abstract{Simple scalar-singlet extensions of the Standard Model with a (spontaneously broken) $Z_2$ symmetry allow for a strong first order electroweak phase transition, as sought in order to realize electroweak baryogenesis. However they generically also lead to the emergence of phenomenologically problematic domain walls. Here we present a framework with a real scalar singlet that features a different thermal history that avoids this problem by never restoring the $Z_2$ symmetry in the early universe. This is accomplished by considering $D>4$ operators that emerge on general grounds, understanding the model as the low energy tail of a more complete theory, like for example in composite Higgs scenarios. Sticking to the latter framework, we present a concrete $SO(6)/SO(5)$ composite realization of the idea. To this end, we additionally provide a complete classification of the structure of the Higgs potential (and the Yukawa couplings) in $SO(6)/SO(5)$ models with fermions in the ${\bf 1, 6, 15}$ or ${\bf 20^\prime}$ of $SO(6)$.
}
\date{\today}

\begin{document}
\maketitle

\section{Introduction}

Understanding how a baryon-asymmetric universe, as we observe it, could have emerged is one of the most important issues in particle physics and cosmology. In fact, this seems to require an extension of the Standard Model (SM) of Particle Physics, which fails to fulfill (quantitatively) two of the three Sakharov criteria~\cite{Sakharov:1967dj} for generating the baryon asymmetry, namely a deviation from thermal equilibrium and the presence of substantial $CP$ violation.

Extensions of the SM with a scalar singlet $\eta$ are promising candidates for baryogenesis at the electroweak scale, by inducing a strong first order electroweak phase transition (EWPhT), providing the out-of-equilibrium situation, and allowing for additional sources of CP violation.
A particularly well studied class of models envisages a real scalar with a $Z_2: \eta\to -\eta$ symmetry (similar to what emerges in non-minimal composite Higgs models~\cite{Gripaios:2009pe,Espinosa:2011eu}) -- which makes $\eta$ contribute to the dark matter abundance, while being protected from collider constraints. Such setups have been shown to allow for a sufficiently strong EWPhT to preserve an adequate baryon asymmetry~\cite{Espinosa:1993bs,Espinosa:2007qk,Profumo:2007wc,Espinosa:2011ax,Espinosa:2011eu,Barger:2011vm,Cline:2012hg,Curtin:2014jma,Kurup:2017dzf,Carena:2018vpt,Carena:2019une}.

A generic challenge in this setting is however the appearance of topological defects, associated to the spontaneous breaking of the $Z_2$ symmetry after $\eta$ acquires a vacuum expectation value (vev) $|\langle \eta \rangle| \equiv v_\eta$ in the thermal evolution of the universe. Patches with $\langle \eta \rangle = +v_\eta$ and $\langle \eta \rangle = -v_\eta$ would get equally populated, which would on the one hand produce potentially dangerous domain walls at the boundaries \cite{Zeldovich:1974uw}, and on the other lead to a cancellation between produced baryon and antibaryon excesses (see, e.g., \cite{Espinosa:2011eu}), requiring additional model building. 

Here we show how a minimal change in the scalar potential can solve these issues via a thermal history in which the $Z_2$ was never a symmetry of the ground state.\footnote{An alternative approach would be to add a small amount of {\it explicit} $Z_2$ breaking~\cite{Espinosa:2011eu,Cline:2021iff}.} In fact, adding higher powers of the scalar-singlet field to the potential can allow for $Z_2$ symmetry non-restoration (SNR) at high temperatures, as we will show below. Such higher dimensional operators are expected to be generated in the presence of new physics addressing other problems of the SM, such as the hierarchy problem or the flavor puzzle.

After having studied the idea in this effective field theory (EFT) extension of the SM in Section~\ref{sec:SNR}, in Section~\ref{sec:SO6} we will provide an explicit realization in the form of a composite Higgs (CH) scenario with $SO(6)/SO(5)$ breaking pattern~\cite{Gripaios:2009pe}, where we will unveil parameter space that had not been considered before. More generally, we will present a comprehensive survey of the Higgs potential in such next-to-minimal CH models~(nMCHMs) for various fermion embeddings and explore their peculiarities, both in general and with respect to the question of generating the sought form of the potential for $Z_2$ SNR. In Section~\ref{sec:Match} we will then match the most promising CH setup to the IR EFT and explore the CH parameter space that leads to a viable SNR. We conclude in Section~\ref{sec:Conc} and provide the Yukawa couplings emerging in the various combinations of $SO(6)$ representations in Appendix~\ref{appx:ferm.allYukawa} for completeness.

\section{$Z_2$ Non-Restoration at High Temperature in Singlet EFT}
\label{sec:SNR}

In the following, we outline a simple Higgs + singlet scenario which exhibits $Z_2$ SNR at high temperature. While present at $T=0$, the $Z_2$ symmetry under which the singlet is odd starts out as broken at high $T$, so as to avoid the formation of topological defects associated with the spontaneous breaking of a discrete symmetry. Therefore, in our envisaged thermal history, the Universe undergoes only one phase transition, which breaks electroweak (EW) symmetry and restores the $Z_2$ symmetry. If strongly first--order (SFO), this EWPhT can fulfill Sakharov's third condition, leading to EW baryogenesis (EWBG), provided there is enough $CP$ violation.

As shown later on, with the given particle content such a scenario cannot be realized at the renormalizable level. However, it can be minimally achieved by extending the renormalizable potential with a dimension--6 singlet--only sextic term.\footnote{We note that higher-dimensional operators have beend studied extensively in scalar-singlet extensions, however not in the context of the phase transition but rather for injecting additional CP violation~\cite{Cline:2012hg,Vaskonen:2016yiu}. Once lifted to an EFT, there is then no reason to not consider other $D=6$ operators (which also appear generically in CH realizations~\cite{Gripaios:2009pe,Espinosa:2011eu}), as done here.} Denoting by $h$ ($\eta$) the background value of the Higgs (singlet) scalar, the $T=0$ tree level potential reads:
\begin{equation}
    V_{\rm tree}(h,\eta) = \frac{\mu_h^2}{2} h^2 + \frac{\lambda_h}{4} h^4 + \frac{\mu_\eta^2}{2} \eta^2 + \frac{\lambda_\eta}{4} \eta^4 + \frac{\lambda_{h \eta}}{2} h^2 \eta^2 +  \frac{\eta^6}{\Lambda^2}.
    \label{eq:potential-tree}
\end{equation}
For the temperature--dependent part of the potential, we work in the high--temperature expansion and retain only the leading $T^2$ contributions:
\begin{equation}
    V_T(h,\eta,T) = \frac{T^2}{2} (c_h h^2 + c_\eta \eta^2) + \frac{5 T^2}{4 \Lambda^2} \eta^4,
    \label{eq:potential-thermal}
\end{equation}
which are added to Eq.~\eqref{eq:potential-tree} to obtain the full potential, denoted as
\begin{equation}
    V(h,\eta,T) \equiv V_{\rm tree}(h,\eta) + V_T(h,\eta,T).
    \label{eq:potential-full}
\end{equation}
Taking into account the leading corrections coming from the top quark and the gauge and scalar sectors, the $c_{h,\eta}$ coefficients are given by:
\begin{equation}
    c_h = \frac{1}{48} \left( 9 g_2^2 + 3 g_1^2 + 12 y_t^2 + 24 \lambda_h + 4 \lambda_{h\eta} \right), \quad c_\eta = \frac{\lambda_{h\eta}}{3} + \frac{\lambda_\eta}{4}.
    \label{eq:ch-ceta}
\end{equation}
With or without the dimension--6 $\eta^6$ term, the necessary and sufficient condition for $Z_2$ SNR at high temperatures is to have a negative $c_\eta$. This way, for sufficiently high $T$, the coefficient of the $\eta^2$ term becomes negative, i.e. $\mu_\eta^2 + c_\eta T^2 < 0$, which destabilizes the origin and sets the global minimum of the potential at $(h,\eta) = (0,w)$. 

In the renormalizable case, corresponding to $\Lambda \to\infty$, one needs $\lambda_\eta > 0$ in order to avoid a runaway direction in the potential, which means that $Z_2$ SNR requires $\lambda_{h\eta} < 0$, cf. Eq.~\eqref{eq:ch-ceta}. At the same time, in order to achieve the desired thermal history of the Universe, we require the coexistence (at intermediate temperatures) of two minima, the $Z_2$--breaking $(0,w)$ and EW minimum $(v,0)$, which become degenerate at the critical temperature $T_c$. The existence of the $(0,w)$ minimum at $T=T_c$ implies that the second derivative of the potential along the $h$ direction is positive at $h=0$, i.e.
\begin{equation}
    V_{hh}(0,w(T_c), T_c) = \mu_h^2 + c_h T_c^2 + \lambda_{h\eta} w(T_c)^2 > 0.
    \label{eq:snr-cdt-1-renormalizable}
\end{equation}
Furthermore, since the potential contains only $h^2$ and $h^4$ terms, the existence of the EW minimum at $T=T_c$ implies that the origin is unstable along the $h$ direction, namely
\begin{equation}
    V_{hh}(0,0, T_c) = \mu_h^2 + c_h T_c^2 < 0.
    \label{eq:snr-cdt-2-renormalizable}
\end{equation}
From the two conditions from Eqs.~\eqref{eq:snr-cdt-1-renormalizable} and \eqref{eq:snr-cdt-2-renormalizable}, it follows that $\lambda_{h\eta} > 0$, which is in contradiction with the SNR condition $\lambda_{h\eta} < 0$. Therefore, as pointed out previously, $Z_2$ SNR cannot be achieved at the renormalizable level in the Higgs + $Z_2$--odd singlet scenario.

Including the $S^6$ term, the conditions from  Eqs.~\eqref{eq:snr-cdt-1-renormalizable} and \eqref{eq:snr-cdt-2-renormalizable} still have to be fulfilled, which means the requirement $\lambda_{h\eta} > 0$ remains. Now, the only way of obtaining $c_\eta < 0 $ is to have a negative $\lambda_\eta$ (as we assume from here on), which becomes viable due to the introduction of the $\eta^6$ term. The role of the latter is to ensure that the potential remains bounded from below~\footnote{Boundedness from below of the scalar potential is ensured as long as $\lambda_h > 0$ and $1/\Lambda^2 > 0$, which always holds in our case.} along the $\eta$ direction, and it turns out to be sufficient for accommodating $Z_2$ SNR. This is the motivation behind considering the $T=0$ potential from Eq.~\eqref{eq:potential-tree}. 

Before performing a numerical analysis of the thermal history induced by the full potential in Eq.~\eqref{eq:potential-full}, we present its main analytical features, namely its ($T$--dependent) critical points. The one at the origin, i.e. at $(h,\eta) = (0,0)$, starts as a saddle point at high--$T$, and later on can turn into a (local) maximum or remain a saddle point. The EW minimum at $(v,0)$ develops as soon as
\begin{equation}
    v^2(T) = -\frac{\mu_h^2 + c_h T^2}{\lambda_h}
    \label{eq:vev[T]}
\end{equation}
becomes positive, leading to a real $v(T)$. Along the $\eta$ direction, the presence of the $\eta^6$ terms leads to the existence of two critical points at $(0,w_\pm)$, with
\begin{equation}
    w_\pm^2(T) = \frac{\Lambda^2}{12} \left( -\lambda_\eta - 5 \frac{T^2}{\Lambda^2} \pm \sqrt{\lambda_\eta^2 - 24 \frac{\mu_\eta^2}{\Lambda^2} + 4 \frac{T^2}{\Lambda^2} \left(\lambda_\eta - 2 \lambda_{h\eta} \right) + 25 \frac{T^4}{\Lambda^4}} \right).
\end{equation}
The critical point at $(0,w_+)$ starts as the $Z_2$--breaking global minimum at high--$T$, and can either remain a minimum till $T=0$ or turn into a saddle point. As long as $w_-(T)^2 > 0$, $(0,w_-)$ exists as a critical point and is either a maximum or a saddle point. Finally, there are two more possible critical points at $(v_{b_-},w_{b_-})$ and $(v_{b_+},w_{b_+})$, with both $v_{b_\pm} \neq 0$ and $w_{b_\pm} \neq 0$. At temperatures where it is real, the former is a saddle point and acts as a barrier between the EW and $Z_2$--breaking minima, rendering the EWPhT strongly first order, whereas the latter develops as a local minimum as soon as $(0,w_+)$ becomes a saddle point. Due to being rather involved, we choose not to show the analytical expressions of $v_{b_\pm}$ and $w_{b_\pm}$.

\begin{figure}
    \centering
    \includegraphics[scale=0.5]{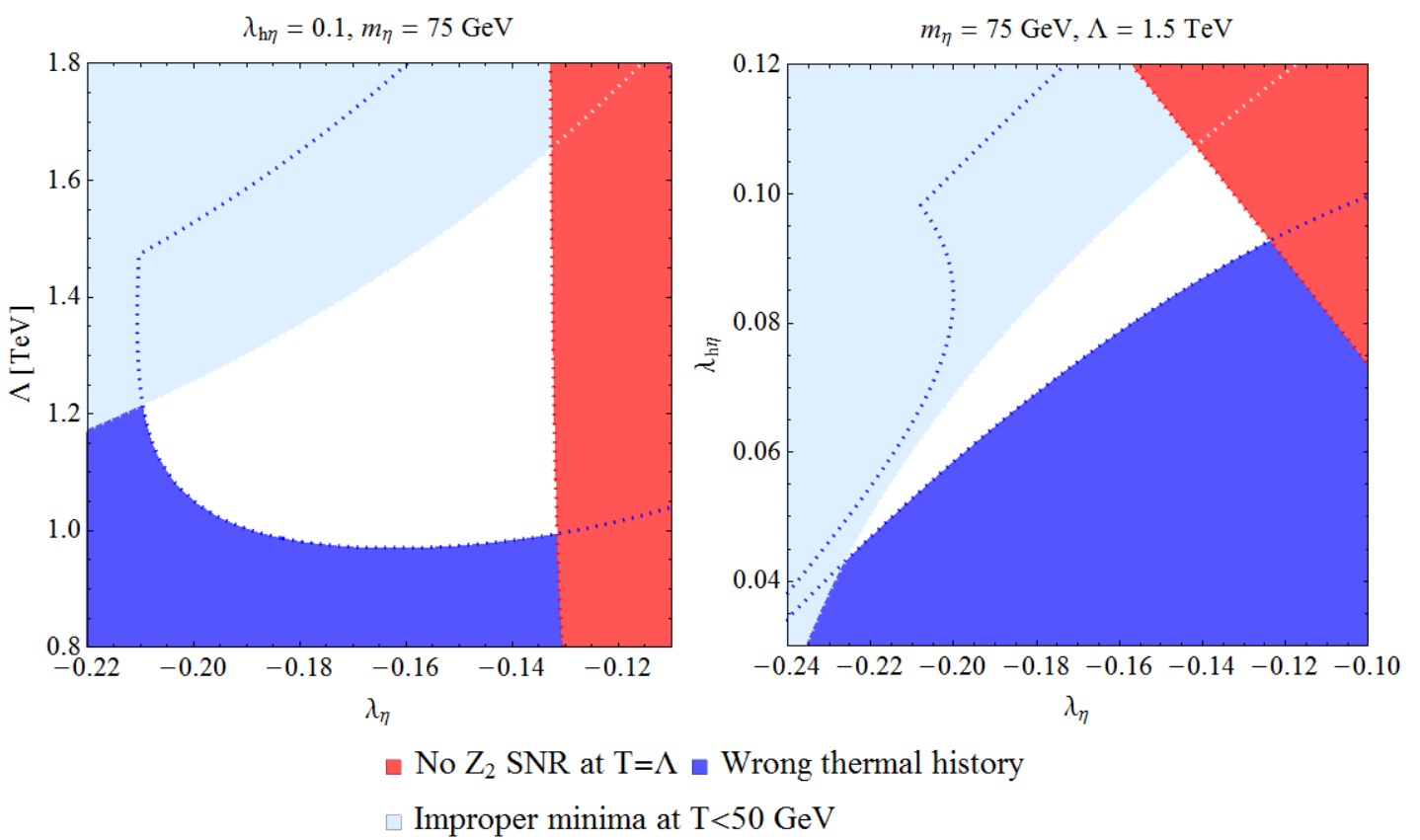} 
    \caption{Visualization of the constraints that we impose in our analysis. The dotted lines represent the boundaries of each of the excluded regions mentioned under the figures. In both panels the singlet mass has been set to $m_\eta = 75$~GeV, whereas in the left (right) panel we fix $\lambda_{h\eta} = 0.1$ ($\Lambda = 1.5$~TeV) and vary $\lambda_\eta$ and $\Lambda$ ($\lambda_\eta$ and $\lambda_{h\eta}$). See text for the explanation of the legend.}
    \label{fig:allowed-regions}
\end{figure} 

\begin{figure}
    \centering
    \hspace*{-10mm}   
    \includegraphics[scale=0.55]{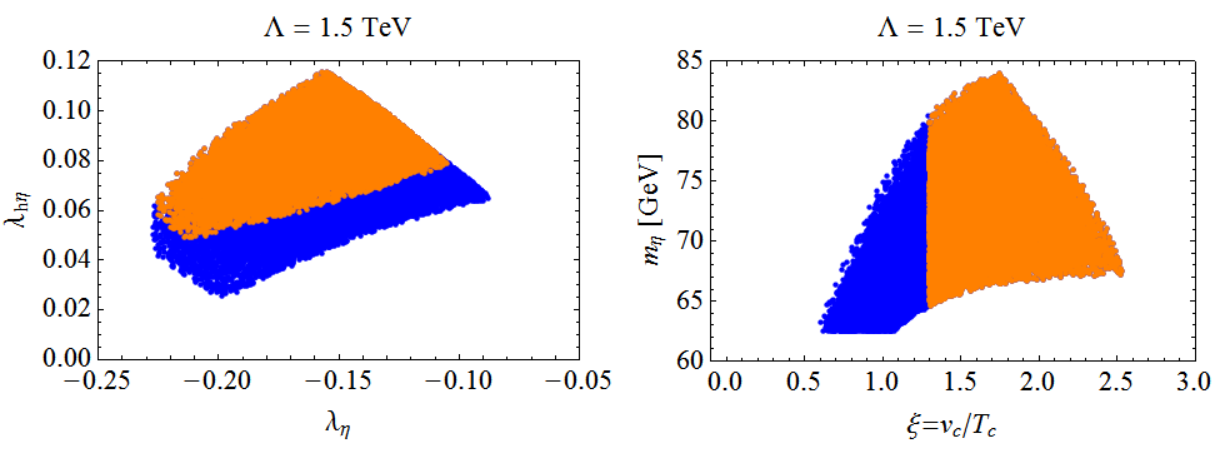} 
    \caption{Values of the parameters for which a one--step (S)FOEWPhT occurs. The orange points correspond to the SFOEWPhT condition, $\xi \geq 1.3$. The cutoff of the dimension--6 operator has been set to $\Lambda=1.5$~TeV.}
    \label{fig:parameter-corr}
\end{figure}

In an initial stage of our numerical analysis, we visualize the various constraints that we impose on our model. We first make sure that, at high $T$, the Universe starts in the $Z_2$--breaking minimum at $(0,w_+)$. For this, we require the origin to be destabilised at $T=\Lambda$ along the $\eta$ direction:
\begin{equation}
    V_{\eta\eta}(0,0,T=\Lambda) = \mu_\eta^2 + c_\eta \Lambda^2 < 0,
    \label{eq:Z2-SNR-cdt}
\end{equation}
which ensures that the origin is not a minimum,\footnote{Alternatively, one can envisage the origin as being a local miminum and $(0,w_+)$ a global minimum at high $T$, with a barrier separating them. However, in such a case, the $Z_2$--breaking minimum remains the global minimum all the way to $T=0$, leaving EW symmetry unbroken, which is clearly an excluded scenario.} leaving $(0,w_+)$ as the only minimum at $T = \Lambda$. We remark that here the initial condition of only populating that $Z_2$-breaking vacuum at high energies and not the opposite $(0,-w_+)$ could emerge dynamically due to inflation blowing up the corresponding patch, making it the full (visible) universe today. Other, disconnected, patches of  $(0,-w_+)$ vacuum would exist, but -- while the full universe would thus be baryon-symmetric -- the local universe would develop an asymmetry. Alternatively, as later on we will embed our model into a composite Higgs scenario, we can assume that the new confining sector is endowed with a mechanism that biases the early Universe towards one of the two equivalent $Z_2$--breaking minima. A more detailed and general analysis of such dynamics is left for future work. Secondly, we impose that the desired thermal history of the Universe is achieved, namely that EWPhT $(0,w_+) \to (v,0)$ is the only PhT in our model, with $T_c$ being the critical temperature at which the two minima become degenerate. Thirdly, in order to make sure that the EWPhT completes, we conservatively require that, below $T=50$~GeV, the only remaining minimum is the EW minimum. While this restriction could in principle over--constrain the parameter space of our model, it will become clear below (e.g. Fig~\ref{fig:parameter-corr}) that imposing it still leads to an untuned and relatively large viable region in parameter space. The goal of the current work is to point out that such regions exist, while a more precise analysis of the boundaries of such regions is beyond the scope of this paper. Lastly, we always choose the singlet mass to be more than half the Higgs mass, $m_\eta > \frac{m_h}{2}$, so as to kinematically close the dangerous $h\to\eta\eta$ decay channel, which is tightly constrained by Higgs decay width measurements~\cite{CMS:2018yfx,ATLAS:2019cid}. Using the CMS $95\%$ CL bound for Higgs decay states, ${\rm BR_{BSM}} < 38\%$~\cite{Workman:2022ynf}, which is the least stringent constraint of its type, and considering the lowest value of the portal coupling from our scans, $\lambda_{h\eta} \simeq 0.025$ (see Fig.~\ref{fig:parameter-corr}), our lower bound for $m_\eta$ gets relaxed to $\sim 61$~GeV from $\frac{m_h}{2} \simeq 62.5$~GeV. The insignificant difference between the two bounds is the reason why we simply choose $m_\eta > \frac{m_h}{2}$ in the following.

The graphical representation of these constraints is shown in Fig.~\ref{fig:allowed-regions}, where in the left (right) panel we fix $\lambda_{h\eta} = 0.1$, $m_\eta = 75$~GeV and vary $\lambda_\eta$ and $\Lambda$ (fix $m_\eta = 75$~GeV, $\Lambda = 1.5$~TeV and vary $\lambda_{h\eta}$ and $\lambda_\eta$). The coloured regions are excluded by the constraints listed above, with the allowed region remaining white. It is interesting to note from the figure that, in our scenario, all the free parameters lie within bounded intervals. Most notably, the value of the scale of New Physics $\Lambda$ is bounded from above, demonstrating the crucial role of the $D=6$ operator.

In a second stage of our numerical analysis, we fix $\Lambda=1.5$~TeV and then scan over the remaining three free parameters, $\lambda_\eta$, $\lambda_{h \eta}$, and $m_\eta$, at the same time imposing the constraints discussed previously. For each viable parameter point, we then calculate the critical temperature $T_c$ corresponding to the $(0,w_+) \to (v,0) $ FOEWPhT, and evaluate the EWPhT strength as $\xi = v(T_c)/T_c $. We show the results of our scan in the scatter plots in Fig.~\ref{fig:parameter-corr} and find that a SFOEWPhT (with $\xi>1.3$), vital for EWBG, can be accommodated. We see that relatively small (absolute) values of the quartic and portal couplings are preferred to arrive at the desired thermal history, as opposed to the $Z_2$ symmetry--restoring scenario from Ref.~\cite{DeCurtis:2019rxl}, which favors ${\cal O}(1)$ values.\footnote{Small portal couplings could also be interesting for potential realizations of dark matter in the singlet-extended SM.}
Qualitatively speaking, the singlet quartic in the $Z_2$--restoring case gets pushed to higher values by imposing that the EW minimum is deeper than the $Z_2$--breaking one at $T=0$: 
\begin{equation}
    -\frac{\mu_\eta^4}{\lambda_\eta} > -\frac{\mu_h^4}{\lambda_h} \, \Rightarrow \, \lambda_\eta > \lambda_h \frac{\mu_\eta^4}{\mu_h^4}.
\end{equation} 
However, imposing the same condition in our $Z_2$ SNR case gives an upper bound on the absolute value of $\lambda_\eta$,
\begin{equation}
    \left| \lambda_\eta \right|^3 < {\rm const} \times \frac{\lambda_h v^4}{\Lambda^4},
\end{equation}
which follows from the depth of the two minima scaling (for negative $\lambda_\eta$) as
\begin{equation}
   - V(v,0,0) \sim \lambda_h v^4, \quad -V(0,w_+,0) \sim  \left| \lambda_\eta \right| w_+^4 \sim  \left| \lambda_\eta \right|^3 \Lambda^4.
\end{equation} 
Furthermore, the singlet is predicted to be rather light, with mass below~$\sim 85$\,GeV. 

\begin{figure}
    \centering
    \hspace*{-10mm}   
    \includegraphics[scale=0.48]{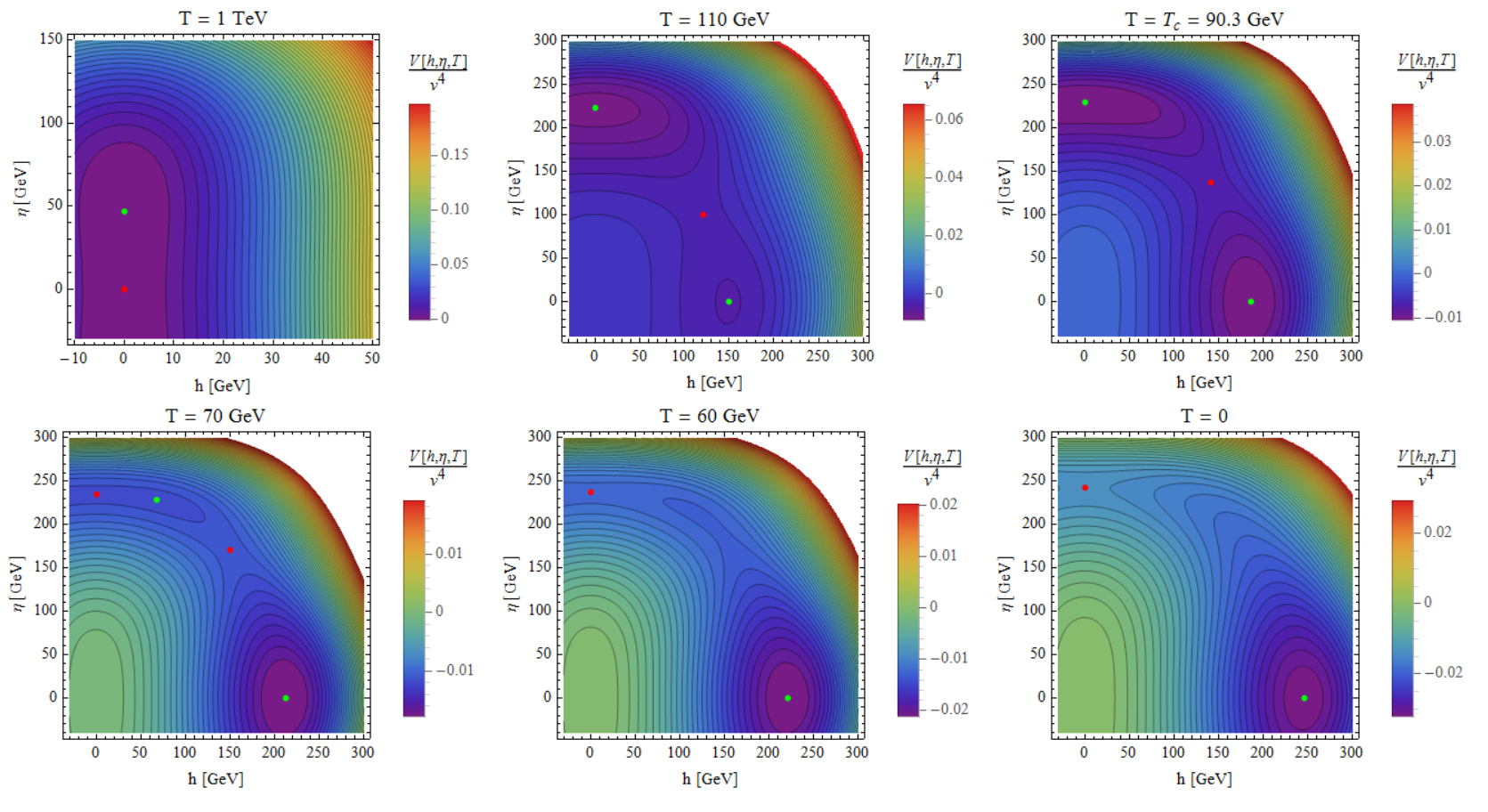} 
    \caption{Thermal evolution of the potential for a benchmark point with $\lambda_\eta = -0.15$, $\lambda_{h \eta} = 0.1$, and $m_\eta = 75$~GeV. The green dots denote local and global minima, whereas the red dots represent saddle points. The cutoff of the dimension--6 operator has been set to $\Lambda=1.5$~TeV.}
    \label{fig:thermal-evolution-potential}
\end{figure}

\begin{figure}
    \centering
    \hspace*{-10mm}   
    \includegraphics[scale=0.38]{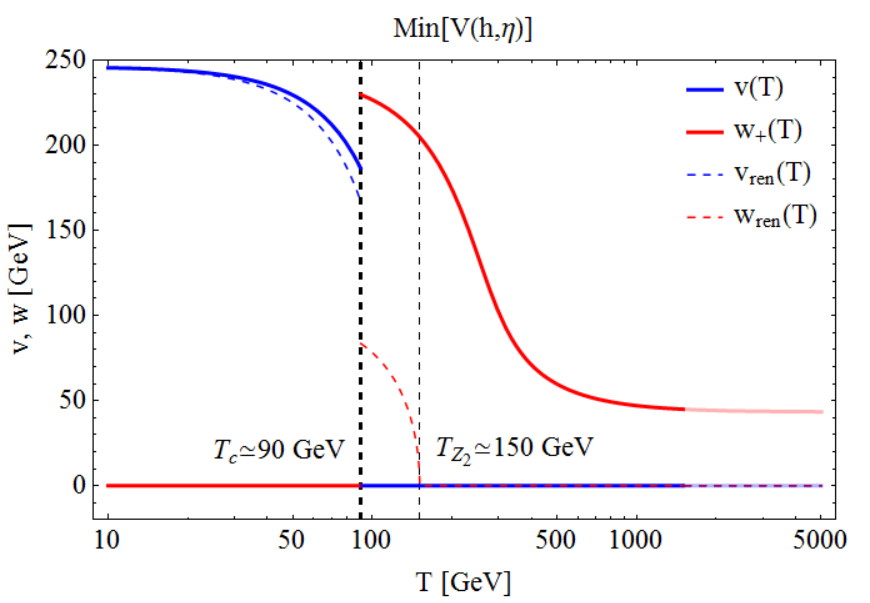} 
    \caption{Thermal evolution of the doublet and singlet vevs in our model for the benchmark point with $\lambda_\eta = -0.15$, $\lambda_{h \eta} = 0.1$, $m_\eta = 75$~GeV, and $\Lambda=1.5$~TeV (solid lines), compared to the conventional case where higher-dimensional operators are neglected (dashed lines), assuming the same critical temperature $T_c \simeq 90\,$GeV. The latter features the characteristic two-step breaking pattern, with $Z_2$ breaking at $T_{Z_2}> T_c$. The region where the validity of the considered EFT starts to break down ($T>\Lambda$) is visualized by faint lines.}
    \label{fig:sum}
\end{figure}

Lastly, in Figs~\ref{fig:thermal-evolution-potential} and~\ref{fig:sum}, we show the thermal history of the Universe for a benchmark point given by $\lambda_\eta = -0.15$, $\lambda_{h \eta} = 0.1$, $m_\eta = 75$~GeV, and $\Lambda=1.5$~TeV. At very high temperatures, the Universe starts in a EW--symmetric phase with broken $Z_2$ symmetry (upper left panel in Fig.~\ref{fig:thermal-evolution-potential}). As the plasma cools down, the EW minimum starts to develop, and is separated from the EW--symmetric minimum by a barrier (upper central panel). Once the critical temperature is attained and the two minima become degenerate (upper right panel), bubbles of the EW--broken phase start to nucleate and the SFOEWPhT proceeds. At a certain temperature below $T_c$, the false vacuum at $(0,w_+)$ turns into a saddle point, and the local minimum at $(v_{b_+},w_{b_+})$ starts to develop~\footnote{This opens up the curious possibility of the minimum at $(v_{b_+},w_{b_+})$ appearing before the $(0,w_+) \to (v,0)$ phase transition completes, which would result in nucleation and collision of bubbles of the broken EW phase in a time--varying background. However, investigating such a scenario is beyond the scope of the present work.} in its place (lower left panel). Later on, at even lower temperatures, the local minimum at $(v_{b_+},w_{b_+})$ and the barrier disappear (lower central panel), leaving the EW minimum as the only minimum all the way to $T=0$ (lower right panel).

The thermal history of our model is plotted and confronted with the standard $Z_2$-restoring case in Fig.~\ref{fig:sum}. Here, we show the $T$-dependent evolution of the doublet and singlet vevs for the mentioned benchmark (solid lines), and compare it to the conventional case where higher-dimensional operators are neglected (dashed lines), assuming the same critical temperature $T_c \simeq 90\,$GeV. The latter features the characteristic two-step breaking pattern, with potentially dangerous $Z_2$ breaking at $T_{Z_2}> T_c$, which can be avoided in the case at hand which features a saturating finite $w_+$ at high temperatures.

Before moving on to the next section, we briefly discuss the impact of including higher--order thermal and Daisy corrections to our analysis. We have randomly selected $20\%$ of the points from our scan and, using them as input, calculated the full one--loop and Daisy corrections to the scalar potential, and checked whether the resulting thermal history of the Universe changes in any way. What we found is that new values of the phase transition strengths $\xi$ deviate by at most $15\%$ from the old ones, while most points only change very little. Moreover, the inclusion of the full thermal corrections in general does not alter the nature of the phase transition, meaning that almost no value of $\xi > 1.3$ gets pushed to $\xi < 1.3$ and vice--versa. This therefore justifies our approach of including only the leading $\mathcal{O}(T^2)$ terms in the scalar potential.


\section{SO(6)/SO(5) Composite Higgs Realization}
\label{sec:SO6}

Here, we present a minimal UV completion of the EFT discussed above in the form of a $SO(6)/SO(5)$ CH model.
The extended scalar sector of this nMCHM~\cite{Gripaios:2009pe} contains in fact a pseudoscalar pseudo Nambu-Goldstone boson (pNGB) singlet~$S$ in addition to the pNGB EW Higgs doublet $H$. Following the partial-compositeness (PC) paradigm \cite{Kaplan:1991dc,Agashe:2004rs,Contino:2003ve,Contino:2006qr}, elementary (SM-like) quarks $q_L$, $q_R$ are coupled to the composite fermionic resonances via linear mixings that explicitly break the global $SO(6)$ symmetry and induce a potential $V(H,S)$ for the scalars, which sensitively depends on the choice of $SO(6)$ representations for the composite matter sector. A main result of this section will be a comprehensive overview of the form of this potential, being a crucial ingredient for EWBG, for the various possible $SO(6)$ representations of the composite fermions in the $SO(6)/SO(5)$ CH, complementing and systematically completing the results available in the literature \cite{Gripaios:2009pe,Chala:2017sjk,DeCurtis:2019rxl,Bian:2019kmg,Xie:2020bkl}.

While EWBG can work in the nMCHM, in the literature so far a specific thermal history with a two-step breaking pattern of $(0,0)\to (0,w) \to (v,0)$ in scalar field space was investigated \cite{Espinosa:2011eu,DeCurtis:2019rxl,Bian:2019kmg,Xie:2020bkl}, realized in a part of the parameter space of the nMCHM. Since, as explained before, such a transition pattern could be problematic, here we will focus on an alternative region, that can allow for the distinct thermal history lined out in Section~\ref{sec:SNR} (which would also be interesting on its own right, beyond the question of emerging topological defects). This space will emerge just by allowing for negative $\lambda_\eta$ and considering $D>4$ operators in the Higgs potential, that automatically appear in the CH framework.

In order to identify setups that feature the structure of couplings envisaged in Section~\ref{sec:SNR}, we employ a spurion analysis where we
formally restore the $SO(6)$-invariance in the linear mixings by uplifting the elementary fermions to transform under the full $SO(6)$ symmetry, even though they actually correspond to incomplete $SO(6)$ multiplets. The spurious symmetry can then be used to constrain the form of the scalar potential.

\subsection{General Setup}
We start by specifying the CH framework.\footnote{For a more complete introduction on $SO(6)/SO(5)$ CH models, see, e.g., \cite{Gripaios:2009pe,Espinosa:2011eu,Redi:2012ha,Niehoff:2016zso,Chala:2017sjk,DeCurtis:2019rxl,Bian:2019kmg,Xie:2020bkl}.}
Schematically, the Lagrangian below the compositeness scale $\Lambda_c$, where the substructure of the scalars would be revealed, reads (after integrating out the heavy resonances)
\begin{equation}
\label{eq:Lag}
    \mathcal{L}_\text{CH} \supset \mathcal{L}^{H,S}_\text{kin} + \mathcal{L}_\text{Yukawa} + {\cal L}_{WZW} - V(H,S) \,.
\end{equation}
Here, $\mathcal{L}^{H,S}_\text{kin}$ denotes the kinetic term for the scalars, $\mathcal{L}_\text{Yukawa}$ contain the light-fermion Yukawa couplings originating from PC, and ${\cal L}_{WZW}$ are Wess-Zumino-Witten (WZW) couplings of the singlet to gauge bosons (which will play no role in the following). The relevant terms, and in particular the Higgs potential $V(H,S)$, will be discussed below.
We are especially interested in $D=6$ corrections to the latter to support the SNR scenario of Section~\ref{sec:SNR}. 
For this purpose, we will analyze several different PC scenarios.
We will concentrate on the top sector $q_L=(t_L,b_L)$, $q_R=t_R$ in particular, but an equivalent analysis can be done for all other quarks. Furthermore, we neglect the gauge boson contribution (which is usually much smaller). 

We denote the 15 generators of $SO(6)$ as $T^A = \{T^{\Bar{A}} , \Hat{T}_6^r\}$, with $T^{\Bar{A}} = \{T_L^a,T_R^a,T_5^i\}$ being the 10 generators of $SO(5)$, where $T_L^a,T_R^a$ correspond to the $SU(2)_L, SU(2)_R$ subgroups, respectively. On the other hand, $\Hat{T}_6^r$ are the 5 broken generators of $SO(6)/SO(5)$, containing a $SU(2)_L \times SU(2)_R$ bi-doublet $(r=1,\dots,4)$ and a singlet $(r=5)$. They can be written as~\cite{Bian:2019kmg,Niehoff:2016zso}
\begin{equation}
\begin{split}
    \big[T^a_L\big]_{IJ} &= -\dfrac{\imath}{2}\left[\dfrac{1}{2} \epsilon^{abc}\left(\delta_{bI}\delta_{cJ}-\delta_{bJ}\delta_{cI}\right)+\left(\delta_{aI}\delta_{4J}-\delta_{aJ}\delta_{4I}\right)\right]
    \\
    \big[T^a_R\big]_{IJ} &= -\dfrac{\imath}{2}\left[\dfrac{1}{2} \epsilon^{abc}\left(\delta_{bI}\delta_{cJ}-\delta_{bJ}\delta_{cI}\right)-\left(\delta_{aI}\delta_{4J}-\delta_{aJ}\delta_{4I}\right)\right]
    \\
    \big[T_5^i\big]_{IJ} &= -\dfrac{\imath}{\sqrt{2}}\left(\delta_{iI}\delta_{5J}-\delta_{iJ}\delta_{5I}\right)
    \\
    \big[\Hat{T}_6^r\big]_{IJ} &= -\dfrac{\imath}{\sqrt{2}}\left(\delta_{rI}\delta_{6J}-\delta_{rJ}\delta_{6I}\right)\,,
\end{split}
\end{equation}
where $a=1,2,3$, $i=1,\dots,4$, $r=1,\dots,5$ and $I,J = 1,\dots,6$. Note that we will need an additional $U(1)_X$ to reproduce the correct hypercharge $Y = X + T_R^3$ for the SM gauge group.
The Goldstone matrix, encoding the dynamics of the pNGBs, is defined as
\begin{equation}
    U\left[\Vec{\Pi}\right] = \exp\left(\imath\dfrac{\sqrt{2}}{f}\Pi_r \Hat{T}_6^r\right)
\label{eq;U.def}
\end{equation}
with $\Pi_r$ the five Goldstone modes and $f = \Lambda_c/(4 \pi)$ the pseudo-Goldstone decay constant. Under $g\in SO(6)$, the Goldstone matrix transforms as (see, e.g., \cite{Panico:2015jxa})
\begin{equation}
    U\left[\Vec{\Pi}\right]\rightarrow g\cdot U\left[\Vec{\Pi}\right] \cdot \hat{h}^T\left[\Vec{\Pi};g\right],\qquad \hat{h}=\left(\begin{array}{cc}
        \hat{h}_5 & 0 \\
         0 & 1 
    \end{array}\right),\qquad \hat{h}_5\in SO(5)\,,
\label{eq;U.transform}    
\end{equation}
providing the non-linear realization of the $SO(6)$-symmetry of the $\Pi$-fields that transform in the fundamental representation of $SO(5)$.
Using the generators $T_L^a$, we can perform a $SU(2)_L$ gauge transformation such that
\begin{align}
    \Vec{\Pi} = \begin{pmatrix}
        \Pi_1, \Pi_2, \Pi_3, \Pi_4, \Pi_5
    \end{pmatrix}^T \overset{SU(2)_L}{\rightarrow}\begin{pmatrix}
        0, 0, 0, \Pi_4, \Pi_5
    \end{pmatrix}^T,
\end{align}
where three Higgs degrees of freedom are eaten by the EW gauge bosons.
However, using this parametrization would lead to a rather involved kinetic term involving trigonometric functions of $\Pi_4$ and $\Pi_5$ so we further redefine~\cite{Bian:2019kmg,Gripaios:2009pe}
\begin{align}
    \frac{h}{f}=\frac{\Pi_{4}}{\sqrt{\Pi_{4}^{2}+\Pi_{5}^{2}}} \sin \frac{\sqrt{\Pi_{4}^{2}+\Pi_{5}^{2}}}{f}, \quad \frac{S}{f}=\frac{\Pi_{5}}{\sqrt{\Pi_{4}^{2}+\Pi_{5}^{2}}} \sin \frac{\sqrt{\Pi_{4}^{2}+\Pi_{5}^{2}}}{f}
\end{align}
which is hereafter simply referred to as unitary gauge.
This leads to the Goldstone matrix
\begin{align}
    U=\left(\begin{array}{cccc}
\mathbb{I}_{3 \times 3} \\
 & 1-\frac{h^{2}}{f^{2}+f \sqrt{f^{2}-h^{2}-S^{2}}} & -\frac{hS}{f^{2}+f \sqrt{f^{2}-h^{2}-S^{2}}} & \frac{h}{f} \\
& -\frac{h S}{f^{2}+f \sqrt{f^{2}-h^{2}-S^{2}}} & 1-\frac{S^{2}}{f^{2}+f \sqrt{f^{2}-h^{2}-S^{2}}} & \frac{S}{f} \\
& -\frac{h}{f} & -\frac{S}{f} & \frac{1}{f} \sqrt{f^{2}-h^{2}-S^{2}}
\end{array}\right)
\end{align}
and the composite sector kinetic term \cite{Chala:2017sjk}
\begin{align}
  \mathcal{L}^{H,S}_\text{kin} \!=  \dfrac{f^2}{4} Tr(d_\mu d^\mu)=(D_\mu H)^\dagger D^\mu H\! +\!\dfrac{1}{2}(\partial_\mu S)^2+\!\dfrac{1}{2f^2}\!\left[\partial_\mu (H^\dagger H)+\dfrac{1}{2}\partial_\mu S^2\right]^2 \!+\! {\cal O}\!\left(\frac{1}{f^4}\!\right)\!\text{,}
\end{align}
where $d_\mu$ is the broken generator part of the Maurer-Cartan form $\omega_\mu = U^{-1} D_\mu U = d_{\mu\, \Bar{A}}\, T^{\Bar{A}}+e_{\mu\, A}\, T^A \equiv d_\mu + e_\mu$, and we rewrote the pNGB Higgs as a complex $SU(2)$ doublet with $H = \frac{1}{\sqrt{2}} (0,h)^T$ in unitary gauge.

\input{FermionEmbeddings}

\section{Matching the EFT to the UV Parameters and Discussion}
\label{sec:Match}

\begin{figure}
    \centering
    \includegraphics[scale=0.49]{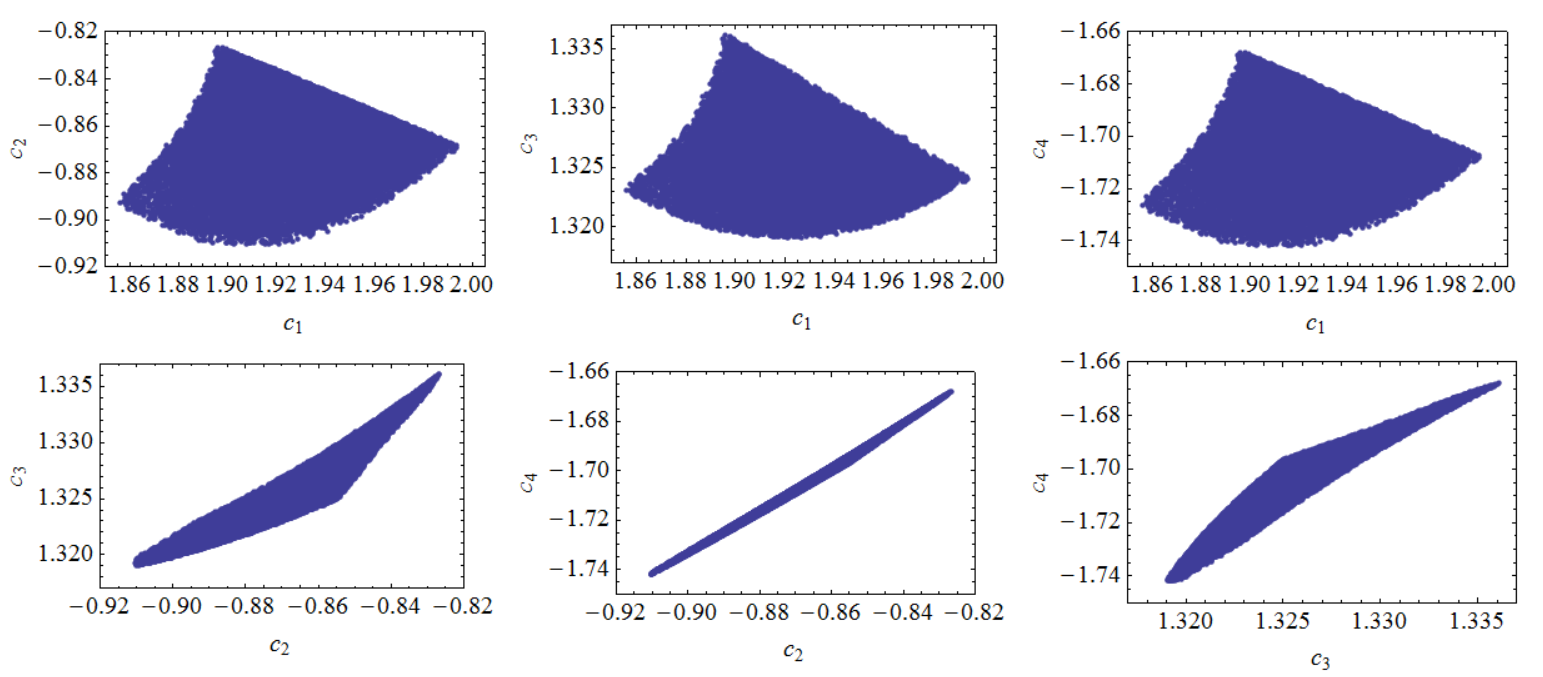} 
    \caption{Values of various UV $c$--parameters that reproduce the IR parameters plotted in Fig.~\ref{fig:parameter-corr}. Here, we have chosen $f=800$~GeV and $g_\ast = 4$.}
    \label{fig:c-parameters}
\end{figure} 

We now turn to matching the parameters appearing in the potential of Eq.~\eqref{eq:potential-tree} to the parameters of the $({\mathbf{20}^\prime_A}_L,\mathbf{20^\prime_\text{AB}}_R)$ model, as given in the bottom section of Table~\ref{tab:FermEmbed.20R}. For simplicity, we set $y_L = y_R = 1$, but the $y_{L,R}$ dependence can be easily restored through simple rescalings (here, for the $D\leq4$ operators, $c_{1,2}$ are always multiplied by $y_L^2$, while $c_{3,4}$ always come with $y_R^2$).

The coefficients of the dimension--6 operators ($h^6$, $h^4 S^2$, $h^2 S^4$, and $S^6$) depend on $\mathcal{O}(10)$ different $c$--parameters, which means we can treat the cutoff $\Lambda$ defined in Eq.~\eqref{eq:potential-tree} basically as a free parameter, within a {\it certain range given by the $y_{L,R}/g_\ast$ scaling}, while the remaining three Wilson coefficients are assumed small enough to not impact our analysis. The smallness of the remaining three Wilson coefficients can be arranged by appropriately choosing the $\mathcal{O}(10)$ different $c$--parameters mentioned above, which we consider as free $\mathcal{O}(1)$ parameters. Even though such a setup would in principle rely on a cancellation/fine--tuning between different $c$'s, the main feature of our setup, $Z_2$ SNR, does not crucially depend on the $h^6$, $h^4 S^2$, and $h^2 S^4$ operators, as they would not affect the potential along the singlet direction, which corresponds to $h=0$. While the $h^4 S^2$ and $h^2 S^4$ terms could in principle play a role at $h \neq 0$ and $S \neq 0$, we expect their effect to be suppressed with respect to the effect of the dimension--4 interaction $h^2 S^2$. Therefore we choose for simplicity to consider only the $S^6$ dimension--6 operator. Setting $\Lambda\sim 1.5$~TeV, see Fig.~\ref{fig:allowed-regions}, we now check to which ${\cal O}(1)$ coefficient $\bar c_4$, appearing in the Lagrangian term $\sim N_c m_\ast^4/16\pi^2 \, (y_{L,R}/g_\ast)^4 \, \bar c_4\, S^6\!/\!f^6$ (see last block of Table~\ref{tab:FermEmbed.20R}), this would correspond. We compare this result to the case of a scaling of $\sim N_c m_\ast^4/16\pi^2\, (y_{L,R}/g_\ast)^6\, \bar c_6 \, S^6\!/\!f^6$, holding for all other models considered.
Plugging in numerical values of $f=800\,$GeV and $g_\ast=4$, we find that $\bar c_4 \sim 10$, which is in reasonable agreement with a (sum of) ${\cal O}(1)$ number(s), while on the other hand $\bar c_6 \sim 200$ would be required. This makes obvious that, while in the latter models the SNR setup is hardly envisagable, it could straightforwardly emerge for the $({\mathbf{20}^\prime_A}_L, \mathbf{20^\prime_\text{AB}}_R)$.

The remaining five IR parameters, namely $\mu_h^2, \, \mu_\eta^2, \, \lambda_h \, \lambda_\eta$, and $\lambda_{h\eta}$, depend on $\cabb$ and four $c$'s, which allows us to express those as functions of the five IR couplings. This exercise gives us an idea of the potential amount of tuning necessary to obtain the potential in Eq.~\eqref{eq:potential-tree} from $\mathcal{O}(1)$ $c$--parameters. We begin by noting that $\cabb$ has a particularly simple expression, depending (weakly) only on the ratio of the singlet and Higgs quartics:
\begin{equation}
    \cabb^2 = \frac{5}{7} \left(1 - \frac{\lambda_\eta}{56 \lambda_h} \right)^{-1}.
\end{equation}
Given the small range of $\lambda_\eta$, cf. Fig~\ref{fig:parameter-corr}, and the weak dependence of $\cabb$ on $\lambda_\eta / \lambda_h$, we find a narrow range for the mixing angle, $\cabb \simeq 0.83 - 0.84$. We have checked that this value is safely within the range of values that can successfully generate the baryon asymmetry in the Universe during the EWPhT~\cite{Espinosa:2011eu} via the CPv-inducing Yukawa term of \eqref{eq:Yuk20}. Even though the full thermal history from Ref.~\cite{Espinosa:2011eu} is different from our thermal history, the $Z_2$--restoring, EW--breaking phase transition, which is responsible for generating the BAU, is common to both setups. Therefore, it is safe to assume that the analysis of Ref.~\cite{Espinosa:2011eu} is applicable to our case.

As for the remaining four $c$--parameters, it turns out the values for the IR couplings can be achieved with natural $O(1)$ values of the UV parameters, as illustrated in Fig.~\ref{fig:c-parameters}, which shows viable regions for various combinations of c's.
The origin of the apparent strong correlation between $c_2$, $c_3$, and $c_4$ can be straightforwardly traced from the expression of the IR parameters for the the $({\mathbf{20}^\prime_A}_L, \mathbf{20^\prime_\text{AB}}_R)$ model. As $\cabb$ can only vary within a narrow interval, fixing the Higgs quartic and mass squared parameter to their SM values basically determines $c_2$ and $c_3$ in terms of $c_4$ (cf. first two rows of the bottom block of Table~\ref{tab:FermEmbed.20R}).  

\section{Conclusions}
\label{sec:Conc}

We explored a new thermal history in the singlet-extended SM with a spontaneously broken $Z_2$ symmetry, considering a general scalar potential. Taking into account higher-dimensional terms in the real singlet, that emerge in UV completions of the setup, we found the possibility of non-restoration of the $Z_2$ symmetry in the early universe. While allowing for a strong first order electroweak phase transition, as required to realize EWBG, the problem of generating domain walls after $Z_2$ breaking (potentially separating patches with a vanishing total baryon number) is avoided.

 After a detailed analysis of this singlet-extended EFT, taking into account various bounds to constrain the parameter space to a well defined viable region, we turned to matching the setup to a motivated UV completion. We found this in the form of a $SO(6)/SO(5)$ nMCHM with a suitable embedding of the SM fermions in the global symmetry.
In particular, we provided a comprehensive overview of the scalar potentials generated in all the different possible realizations of the fermion sector, which is also relevant for the question of a viable EWSB (and scalar phenomenology) in nMCHMs in general. Moreover, we provided explicit expressions for the emerging Yukawa couplings in the different models, including CPv terms involving the Higgs and the singlet, that allow to fulfill all Sakarov criteria for a viable baryogenesis. 

Finally, identifying the $(\mathbf{20^\prime_{\text{A}}}_L,\mathbf{20}^\prime_R)$ variant as the most promising model, we presented a numerical analysis and showed that the $Z_2$ SNR setup emerges with natural values of the model coefficients, matching all criteria identified in the EFT analysis.
It would be interesting to explore if the found thermal history could also emerge in other motivated models beyond the SM, like in the recently proposed model of $SU(6)$ Gauge-Higgs Grand Unification with an extra scalar singlet~\cite{Angelescu:2021nbp,Angelescu:2021qbr}, or in further models with extended scalar sectors.

\section*{Acknowledgments}

We are grateful to Arthur Hebecker for useful discussions.

\appendix
\input{Appendix}

\bibliographystyle{hunsrt.bst}    
\bibliography{BG_CH}

\end{document}

%% file: FermionEmbeddings.tex
\newcommand{\sth}[2][2]{\sin^{#1}{\theta_\text{#2}}\,}
\newcommand{\cth}[2][]{\cos^{#1}{2\theta_\text{#2}}\,}
\newcommand{\const}[1]{c_\text{#1}}
\newcommand{\sabb}{\mathfrak{s}_\theta}
\newcommand{\cabb}{\mathfrak{c}_\theta}
\newcommand{\cabbs}{\mathfrak{c}_{2\theta}}\,
\newcommand{\call}[1]{\mathfrak{c}_{\theta \text{#1}}}
\newcommand{\sall}[1]{\mathfrak{s}_{\theta \text{#1}}}
\newcommand{\Lyuk}{\mathcal{L}_\text{Yukawa}}
\subsection{Fermion Embeddings and Scalar Potential}\label{sec:ferm.embeds}
Regarding the fermions, the simplest option for the right-handed SM quarks is to embed $t_R$ in a singlet of $SO(6)$ (being either partially or fully composite), for which the potential only receives contributions from the spurions corresponding to the left-handed SM quarks, $q_L$, see below. Concerning higher representations, the spinorial $\mathbf{4}$ is not considered in general, as it does not obey custodial symmetry for the $Z\bar b b$ couplings~\cite{DeCurtis:2019rxl}. On the other hand $\mathbf{10}$-representations are neglected because they fail to produce a singlet potential as they do not break the $SO(2)_S \subset SO(6)$ subgroup \cite{DeCurtis:2019rxl}, while the  $\mathbf{20}$ and $\mathbf{20^{\prime\prime}}$ representations do not yield valid SM quark embeddings at all. Therefore, we consider combinations of $q_L \in (\mathbf{6},\mathbf{15},\mathbf{20^\prime})$ and $t_R \in (\mathbf{1}, \mathbf{6},\mathbf{15},\mathbf{20}^\prime)$~\cite{Bian:2019kmg,Xie:2020bkl,DeCurtis:2019rxl}.

The $\mathbf{6}$ decomposes under $SO(6)\times U(1)_X\rightarrow SO(5)\times U(1)_X\rightarrow SO(4)\times U(1)_X\rightarrow SU(2)\times U(1)_Y$ as
\begin{align*}
    \mathbf{6}_{2/3} &\rightarrow \mathbf{5}_{2/3}\oplus \mathbf{1}_{2/3} \\
    &\rightarrow \left[\mathbf{4}_{2/3}\oplus \mathbf{1}_{2/3}\right] \oplus \mathbf{1}_{2/3} \\ 
    &\rightarrow \left[\left(\mathbf{2}_{7/6}\oplus \mathbf{2}_{1/6}\right)\oplus \mathbf{1}_{2/3}\right] \oplus \mathbf{1}_{2/3}.
\end{align*}
As there is only one $\mathbf{2}_{1/6}$, the $Q_L$ embedding in $SO(6)$ is unique, while $t_R$ resides in a superposition of the two $\mathbf{1}_{2/3}$, parameterized by the angle $\theta_{6R}$:
\begin{align}
    Q_L^6 = \dfrac{1}{\sqrt{2}} \begin{pmatrix} \left(Q_L^4\right)^T & 0 & 0 \end{pmatrix}^T,\quad t_R^6 = \begin{pmatrix} 0 & 0 & 0 & 0 & t_R e^{\imath\phi_{6R}} \cos{\theta_{6R}}& t_R \sin{\theta_{6R}} \end{pmatrix}^T\,,
\end{align}
where $Q_L^4 = \begin{pmatrix} \imath b_L & b_L &\imath t_L & -t_L \end{pmatrix}^T $.
In the following we employ $\phi_{6R} = \pm \pi/2$, in agreement with a CP-conserving top coupling~\cite{Redi:2012ha}. For our purposes, we want to avoid $\theta_{6R} = \pm \pi/4$, for which the singlet stays a pure Goldstone. The angle $\theta_{6R} = \pm \pi/2$ on the other hand yields a $\mathrm{Z}_2$ symmetry with the singlet being odd, making it a dark matter candidate~\cite{Redi:2012ha}, but we will not explore this direction further here.

The $\mathbf{15}$ decomposes as
\begin{align*}
    \mathbf{15}_{2/3} &\rightarrow \mathbf{10}_{2/3} \oplus \mathbf{5}_{2/3} \\
    &\rightarrow \left[\mathbf{3}_{2/3}\oplus \mathbf{3}_{2/3}^\prime\oplus \mathbf{4}_{2/3}\right]\oplus \left[\mathbf{4}_{2/3}\oplus \mathbf{1}_{2/3}\right]\nonumber\\
    &\rightarrow \left[\mathbf{3}_{2/3}\oplus \left(\mathbf{1}_{5/3}\oplus \mathbf{1}_{2/3}\oplus \mathbf{1}_{-1/3}\right)\oplus \left(\mathbf{2}_{7/6}\oplus \mathbf{2}_{1/6}\right) \right]\oplus \left[\left(\mathbf{2}_{7/6}\oplus \mathbf{2}_{1/6}\right)\oplus \mathbf{1}_{2/3}\right].
\end{align*}
Thus, $q_L$ can be embedded in the $\mathbf{10}$ (A) or the $\mathbf{5}$ (B) of $SO(5)$,
\begin{align}
    Q_L^{15_\text{A}} = (Q_L^4)_j T_5^j, \quad Q_L^{15_\text{B}} = \imath (Q_L^4)_j \Hat{T}_6^j\,.
\end{align}
A general embedding would be given by
\begin{align}
    Q_L^{15} = \cos \theta_{15L} e^{\imath \phi_{15L}}Q_L^{15_\text{A}}+\sin \theta_{15L} Q_L^{15_\text{B}}\,,
\end{align}
however, since $\mathbf{15}_\text{B}$ is heavily constrained by $Zb\bar{b}$ couplings~\cite{Bian:2019kmg} it is dropped below, corresponding to $\theta_{15L}=0$.
Similarly, the $t_R$ can be embedded as 
\begin{align}
    t_R^{15} =  \cos{\theta_{15R}} e^{\imath \phi_{15R}} T_R^3 t_R +\sin{\theta_{15R}} \Hat{T}_6^5 t_R .
\end{align}

Finally, the $\mathbf{20}^\prime$ decomposes as
\begin{align*}
    \mathbf{20}_{2 / 3}^{\prime} & \rightarrow \mathbf{14}_{2 / 3} \oplus \mathbf{5}_{2 / 3} \oplus \mathbf{1}_{2 / 3} \\
    & \rightarrow\left[\mathbf{9}_{2 / 3} \oplus \mathbf{4}_{2 / 3} \oplus \mathbf{1}_{2 / 3}\right] \oplus\left[\mathbf{4}_{2 / 3} \oplus \mathbf{1}_{2 / 3}\right] \oplus \mathbf{1}_{2 / 3} \\
    & \rightarrow\left[\left(\mathbf{3}_{5 / 3} \oplus \mathbf{3}_{2 / 3} \oplus \mathbf{3}_{-1 / 3}\right)\! \oplus\!\left(\mathbf{2}_{7 / 6} \oplus \mathbf{2}_{1 / 6}\right) \!\oplus\! \mathbf{1}_{2 / 3}\right] \oplus\left[\left(\mathbf{2}_{7 / 6} \oplus \mathbf{2}_{1 / 6}\right) \!\oplus\! \mathbf{1}_{2 / 3}\right] \oplus \mathbf{1}_{2 / 3}\,,
\end{align*}
and we can write $Q_L^{20^\prime}$ as a superposition of the embeddings in a $\mathbf{14}$ (A) and a $\mathbf{5}$ (B) of SO(5),
\begin{align}
    Q_L^{20^\prime_\text{A}} = \dfrac{1}{2}
    \begin{pmatrix}
        0_{4\times4}& Q_L^4 & 0_{4\times1}\\
        \left(Q_L^4\right)^T & 0 & 0\\
        0_{1\times4} & 0 & 0
    \end{pmatrix},
    \quad
        Q_L^{20^\prime_\text{B}} = \dfrac{1}{2}
    \begin{pmatrix}
        0_{4\times4} & 0_{4\times1}& Q_L^4\\
        0_{1\times4} & 0 & 0\\
        \left(Q_L^4\right)^T & 0 & 0
    \end{pmatrix}\,,
\end{align}
with a general realization given by 
\begin{align}
\label{eq:20}
    Q_L^{20^\prime} = \cos \theta_{20L} e^{\imath \phi_{20L}}Q_L^{20^\prime_\text{A}}+\sin \theta_{20L} Q_L^{20^\prime_\text{B}}.
\end{align}
While the $\mathbf{20^\prime_\text{B}}$ leads again to large corrections to $Zb\bar{b}$~\cite{Xie:2020bkl}, for most models considered, sticking to $Q_L^{20^\prime_\text{A}}$ would not lead to a phenomenologically viable model and we will keep the superposition, see below. Finally, $t_R$ can be embedded in a superposition of the $\mathbf{14}$ (A), $\mathbf{5}$ (B) and $\mathbf{1}$ (C) representations,
\begin{align}
    t_R^{20^\prime_\text{A}} &= \dfrac{1}{2\sqrt{5}}
    \begin{pmatrix}
        -\mathbb{I}_{4\times4} t_R  & 0_{4\times2}\\
        0_{2\times4} & 2(\mathbb{I}_{2\times2}+\sigma^3)t_R
    \end{pmatrix},
    \quad
        t_R^{20^\prime_\text{B}} = \dfrac{1}{\sqrt{2}}
    \begin{pmatrix}
        0_{4\times4} & 0_{4\times2}\\
        0_{2\times4} & \sigma^1 t_R
    \end{pmatrix},\nonumber \\
    t_R^{20^\prime_\text{C}} &= \dfrac{1}{\sqrt{30}}
    \begin{pmatrix}
        -\mathbb{I}_{5\times5} t_R  & 0_{5\times1}\\
        0_{1\times5} & 5 t_R
    \end{pmatrix},
\end{align}
where $\sigma^a$ are the Pauli matrices, leading to
\begin{align}
    t_R^{20^\prime} =  \cos{\theta_{20R1}}e^{\imath \phi_{20R1}}t_R^{20^\prime_A}+\sin{\theta_{20R1}}\cos{\theta_{20R2}}e^{\imath \phi_{20R2}}t_R^{20^\prime_B}+\sin{\theta_{20R1}}\sin{\theta_{20R2}}t_R^{20^\prime
    _C}.
\end{align}
\input{MinExtSILH}
\subsubsection{Survey of Different Embeddings}

In this section we will follow the procedure of Section \ref{sec:MinSILH} to calculate the potential for various fermion embeddings, aiming to carve out differences in the hierarchies of couplings.  Note that in the general case we get contributions from the $t_R$-spurions, too. We will thus distinguish between $y_L$ and $y_R$-type spurions with
\begin{align}
    y_t \propto \dfrac{y_R y_L}{g^*},
\end{align}
as per the definition of $\mathcal{L}_\text{Yukawa}$, see Appendix \ref{appx:ferm.allYukawa}, where the Yukawa terms for the realizations chosen below are given. 

In \autoref{tab:FermEmbed.1R}-\ref{tab:FermEmbed.20R}
we provide an overview of the spurion-order at which the different terms in the potential appear for various viable combinations of $q_L$ and $t_R$ embeddings, as well as the leading contributions to each term. We then compactly summarize this hierarchy in \autoref{Tab:SO6.spurion-overview}. We only take into account embeddings featuring no explicit CP-breaking, i.e. that only generate even powers of the pseudoscalar $S$ in the potential, which basically fixes our choices for the angles $\phi_i$ below. In $\Lyuk$, a non-zero top mass as well as a CP-violation (CPv) inducing term $\propto \imath S h \bar{t}\gamma^5 t$ need to arise, the latter due to the last Sakharov criterion, constraining the viable embeddings further.\footnote{In the EFT of Section~\ref{sec:SNR}, with $\eta$ a pseudoscalar, the necessary CP violation could be injected via a ($D\!=\!6$) $\imath S^ 2 h \bar{t}\gamma^5 t$ operator \cite{Cline:2012hg}. We note that, even though the CH model features in general only a (spontaneously broken) CP symmetry and no additional $Z_2$, regarding the discussed thermal evolution both setups are equivalent.} To minimize corrections to $Zb\Bar{b}$, we generally restrict ourselves to ${\mathbf{15}_{\!\text{A}}}_L$ ($\theta_{15L}= 0$) and ${\mathbf{20}^\prime_{\!\text{A}}}_L$ ($\theta_{20L}= 0$) and assume $\langle S \rangle|_{T=0} = 0 $ (c.f.~\cite{Bian:2019kmg,Xie:2020bkl}). Only for the $({\mathbf{20}^\prime_{\!\text{A}}}_L,{\mathbf{1}}_R)$ model, $\theta_{20L}\neq 0$ is strictly needed to generate a top mass term.
Note that wherever we display only a limit $\theta_{L/R} = 0$, going to finite values of the angle does not change the hierarchy of the terms, except for the $({\bf 20}^\prime_L,{\bf 6}_R)$ and $({\bf 20}^\prime_L,{\bf 15}_R)$, where the Higgs quartic and sextic terms appear at lower orders, see below. Also note that no $S$-potential arises for combinations of $Q_L \in\left\{\mathbf{6},\mathbf{15}_{\theta_{15L} = \pi/4}, \mathbf{20}_{\theta_{20L} = \pi/4}\right\}$ with $t_R \in\left\{\mathbf{1},\mathbf{6}_{\theta_{6R} = \pi/4},\mathbf{15}\right\}$, for which the singlet shift symmetry is conserved. Finally, we will assess whether the respective models fit the SNR scenario of Section \ref{sec:SNR}. 

We start in \autoref{tab:FermEmbed.1R} with the models where the $t_R$ is realized as a composite singlet. Here, only $Q_L$ in a $\mathbf{20}^\prime_L$ yields a viable setup, with the results already discussed in Section~\ref{sec:MinSILH}. They are summarized here again for completeness with a slightly different ordering, using $\phi_{20L} = -\pi/2$, while $\theta_{20L}\neq \{0,\dfrac{\pi}{2}\}$ ensures a nonzero top-quark mass and the presence of CPv (and similarly for the following tables). For the ${\bf 6}_L$, no singlet potential is generated as the embedding does not break the singlet shift symmetry, whereas for the ${\bf15}_L$ the top quark remains massless. For the $(\mathbf{20}^\prime_L,\mathbf{1}_R)$ model, as discussed before, the singlet sextic (quartic) interaction is generated at order $(y_L/g_\ast)^6$ ($(y_L/g_\ast)^4$), making it challenging to realize the $Z_2$ SNR scenario with natural ${\cal O}(1)$ dimensionless coefficients. We finally note that both the renormalizable scalar portal as well as the pure ($D\leq 4$) Higgs terms arise at LO in spurions.

Next, we consider the models with a ${\bf 6}_R$ embedding in \autoref{tab:FermEmbed.6R}. For the $({\bf 6}_L,{\bf 6}_R)$ case, $\phi_{6R} = -\pi/2$ ensures a CP-conserving potential. Despite the fact that no CPv inducing term arises for $({\bf 15}_L,{\bf 6}_R)$, we still write down (in gray font) the model for $\theta_\text{15L}=0$, and $\phi_{6R}-\phi_{15L}= \pi/2$. For the $({\bf 20^\prime}_L,{\bf 6}_R)$ we choose $\phi_{6R} = \phi_{20L} = -\pi/2$.
As mentioned above, for $\theta_{6R} =\pi/4$ and $\theta_{20L} =\pi/4$, the singlet becomes a true Goldstone boson in both the $(\mathbf{20^\prime}_L,\mathbf{6}_R)$ model and the $(\mathbf{6}_L,\mathbf{6}_R)$.
For all other cases, as before, the $S^n$ terms arise at the $n$th order in spurions. The renormalizable portal, on the other hand, emerges only at NLO in spurions, besides for the $({\bf 20}^\prime_L,{\bf 6}_R)$. More importantly, the Higgs quartic coupling also emerges only at NLO - both couplings are thus generically suppressed, while the Higgs mass term arises at LO. In particular this latter hierarchy leads to a larger fine-tuning of the model, see, e.g., \cite{Panico:2012uw,Matsedonskyi:2012ym,Carmona:2014iwa}. Only when going to finite $\theta_{20L}\neq 0$ in the $({\bf 20}^\prime_L,{\bf 6}_R)$ model, the Higgs quartic (sextic) interaction is generated at at LO (NLO), fitting the hierarchies in the $(\mathbf{ 20}^\prime_L,\mathbf{ 1}_R)$ model. 

The models with a ${\bf 15}_R$ are displayed in \autoref{tab:FermEmbed.15R}. The singlet shift symmetry is unbroken for the $({\bf 6}_L,{\bf 15}_R)$ model, as well as for $\theta_{15(20)L} = \pi/4$ in the other two cases. Due to the $Zbb$ coupling constraints, the latter choice of angles is not preferred in the first place and we choose
 $({\bf 15_\text{A}}_L,{\bf 15}_R)$ with $\phi_{15L}=\phi_{15R}=0$ and $({\bf 20^\prime_\text{A}}_L,{\bf 15}_R)$ with $\phi_{20L}=-\pi/2, \phi_{15R}=-\pi$, which fulfill all basic conditions. 
Once more, the $S^4 (S^6)$ terms arise only at NLO (NNLO) in spurions. As before, when going to finite $\theta_{20L}\neq 0$, both the Higgs mass and quartic coupling arise at the same order.

We finally move to the $\mathbf{20}^\prime_R$ models, displayed in \autoref{tab:FermEmbed.20R}, which will turn out to be the most interesting for us. Out of all the models, the one that fits the SNR scenario best is the $(\mathbf{20^\prime_{\text{A}}}_L,\mathbf{20}^\prime_R)$. Here, while we need a $\mathbf{20^\prime_\text{B}}_R$ contribution to generate the top mass, determined by $y_t \propto y_L y_R^\ast \sin{\theta_{20R1}}$, and $\mathbf{20^\prime_\text{A}}_R$ to generate the CPv-inducing operator, an additional $\mathbf{20^\prime_\text{C}}_R$ contribution would not change the order of operators (yet modify their correlation, see below). Accordingly, we initially focus on $\theta_{20L}=0$, $  \theta_{20R2}=0$, $\phi_{20R1} = \pi/2$, $ \phi_{20R2} = 0$.

The same considerations lead to focusing on the $(\mathbf{15_{\text{A}}}_L,\mathbf{20^\prime_\text{AB}}_R)$ setup, while for the $(\mathbf{6}_L,\mathbf{20^\prime}_R)$, both $\mathbf{20^\prime_\text{AB}}_R$ and $\mathbf{20^\prime_\text{BC}}_R$ would work equally fine. Therefore, we display only $\theta_{20R2}=0$ models in the table. For the Yukawa terms we employ $\phi_{20R1}-\phi_{15L}=\pi$, $ \phi_{20R2}-\phi_{15L}=\pi/2$ ($\phi_{20R1}=\pi$,  $\phi_{20R2}=-\pi/2$) in the former (latter) case. Note that here, the $\mathbf{6}_L$ and $\mathbf{15_{\text{A}}}_L$ models predict opposite signs for the $h^4$ and $h^2 S^2$ terms and thus do not fulfill the conditions for the $Z_2$ SNR scenario of Section~\ref{sec:SNR}.

This can be remedied by allowing nonzero $\theta_{R2}$, where we can fulfill the constraints on signs and magnitude of the couplings for points within $\theta_{20R2}\in \left(0.70, 1.50\right) \cup \left(1.64, 2.45\right)$ for $\theta_{20R1}\in \left(1.11, 1.50\right)\cup \left(1.11, 1.50\right)\!+\!\pi$ and $\theta_{20R2}\in \left(0.70, 1.50\right)\!+\!\pi \cup \left(1.64, 2.45\right)\!+\!\pi$ for $\theta_{20R1}\in \left(0.70, 1.11\right) \cup \left(0.70, 1.11\right)\!+\!\pi$.

Coming back to our model of choice, the $({\mathbf{20}^\prime_A}_L, \mathbf{20^\prime_\text{AB}}_R)$, we can easily create the sought pattern of couplings to realize the $Z_2$ SNR scenario, where the $S^6$ ($S^4$) terms arise already at NLO (LO) in spurions, which allows for non-negligible contributions.
Also the fact that the Higgs quartic coupling arises at LO in spurions is interesting regarding the naturalness of the setup.
We will provide a more quantitative evaluation of this model in the next section.
\include{EmbedsTables}
\begin{table}[h]
  \centering
  \caption[Overview of $V(h,S)$ terms arising at $n$th order in spurions]{Overview of $V(h,S)$ terms arising at $n$th order in spurions, with brackets denoting special cases of embeddings described in the text (not noting the cases for which $S$ remains a Goldstone boson), i.e., in the $\mathbf{20^\prime}_{L}$ block, $\theta_\text{20L}= 0$ in the first column, $\theta_\text{20L}\neq 0 $ for the second and third and $\theta_\text{20R1} = 0 $ for the $(\mathbf{20^\prime_{\text{A}}}_L,\mathbf{20}^\prime_R)$.}\label{Tab:SO6.spurion-overview}
  \begin{tabular}{|c|c|l|l|l|l|}
      \hline
      \multicolumn{2}{|c|}{}&$\mathbf{1}_R$&$\mathbf{6}_R$&$\mathbf{15}_R$ & $\mathbf{20^\prime}_R$\\ 
      \hline
      \multirow{9}{*}{$\mathbf{6}_L$}
      & $h^2$&\multirow{9}{*}{\parbox{1.1cm}{\vspace{-30pt}\centering\color{gray} No\hspace{12pt} S-pot. \color{black}} }& $2^\text{nd}$ &\multirow{9}{*}{\parbox{1.1cm}{\vspace{-30pt}\centering\color{gray} No\hspace{12pt} S-pot. \color{black}} } & $2^\text{nd}$ \\
      & $h^4$& & $4^\text{th}$ & & $2^\text{nd}$\\
      & $h^6$& & $6^\text{th}$ & & $4^\text{th}$\\
      & $S^2$& & $2^\text{nd}$ & & $2^\text{nd}$\\
      & $S^4$& & $4^\text{th}$ & & $2^\text{nd}$\\
      & $S^6$& & $6^\text{th}$ & & $4^\text{th}$\\
      & $h^2 S^2$& & $2^\text{nd}$ & & $2^\text{nd}$\\
      & $h^2 S^4$& & $4^\text{th}$ & & $4^\text{th}$\\
      & $h^4 S^2$& & $6^\text{th}$ & & $4^\text{th}$\\
      \hline
      \multirow{9}{*}{$\mathbf{15_\text{A}}_L$}
      & $h^2$&\multirow{9}{*}{\parbox{1.1cm}{\vspace{-30pt}\centering \color{gray} No top mass \color{black}} } & \multirow{9}{*}{\parbox{1.1cm}{\vspace{-30pt}\centering\color{gray} No\hspace{12pt} CPV \color{black}} } & $2^\text{nd}$& $2^\text{nd}$\\
      & $h^4$& &  & $4^\text{th}$ & $2^\text{nd}$\\
      & $h^6$& &  & $6^\text{th}$ & $4^\text{th}$\\
      & $S^2$& &  & $2^\text{nd}$ & $2^\text{nd}$ \\
      & $S^4$& &  & $4^\text{th}$ & $2^\text{nd}$\\
      & $S^6$& &  & $6^\text{th}$ & $4^\text{th}$\\
      & $h^2 S^2$& & & $4^\text{th}$ & $2^\text{nd}$\\
      & $h^2 S^4$& & & $6^\text{th}$ & $4^\text{th}$\\
      & $h^4 S^2$& & & $6^\text{th}$ & $4^\text{th}$\\
      \hline
      \multirow{9}{*}{$\mathbf{20}^\prime_L$}
      & $h^2$& $2^\text{nd}$ & $2^\text{nd}$ & $2^\text{nd}$ & $2^\text{nd}$\\
      & $h^4$& $2^\text{nd}$\,$(4^\text{th})$ & $4^\text{th}$\,$(2^\text{nd})$ & $4^\text{th}$\,$(2^\text{nd})$ & $2^\text{nd}$\,$(4^\text{th})$\\
      & $h^6$& $4^\text{th}$\,$(6^\text{th})$ & $6^\text{th}$\,$(4^\text{th})$ & $6^\text{th}$\,$(4^\text{th})$ & $4^\text{th}$\,$(6^\text{th})$\\
      & $S^2$& $2^\text{nd}$ & $2^\text{nd}$ & $2^\text{nd}$ & $2^\text{nd}$ \\
      & $S^4$& $4^\text{th}$ & $4^\text{th}$ & $4^\text{th}$ & $2^\text{nd}$\\
      & $S^6$& $6^\text{th}$ & $6^\text{th}$ &$6^\text{th}$ &$4^\text{th}$\\
      & $h^2 S^2$& $2^\text{nd}$ & $2^\text{nd}$& $2^\text{nd}$ & $2^\text{nd}$\\
      & $h^2 S^4$& $4^\text{th}$ & $4^\text{th}$ & $4^\text{th}$ & $4^\text{th}$\\
      & $h^4 S^2$& $4^\text{th}$ & $4^\text{th}$ & $4^\text{th}$ & $4^\text{th}$\\
      \hline
  \end{tabular}
\end{table}

%% file: MinExtSILH.tex
\subsubsection{Warm-Up: ($\mathbf{20^\prime,1}$) Minimally Extended SILH}\label{sec:MinSILH}
As a first explicit example, we will evaluate the fermion couplings and the pNGB potential for the
$(\mathbf{20^\prime},\mathbf{1})$ model with the $Q_L$ in a $\mathbf{20^\prime}$ of $SO(6)$ and a fully composite $t_R$ singlet, allowing a viable EW symmetry breaking (EWSB) from leading (LO) order in the PC expansion~\cite{Chala:2017sjk}. Compared to the analyses available in the literature, here and below we will take into account higher orders in the scalar potential, which are crucial to reveal regions of parameter space allowing for the novel SNR thermal history.
While we will follow the parametrization given above, the analysis is consistent with \cite{Chala:2017sjk}, which we will further comment on later.
We can rewrite the general embedding in a ${\bf 20^\prime}$ of Eq.~\eqref{eq:20} as
\begin{align}
Q_L^{20^{\prime}}&=\cos \theta_{20L} e^{i \phi_{20L}}  Q_L^{20_{A}^{\prime}} + \sin \theta_{20L} Q_L^{20_{B}^{\prime}} \nonumber\\
&=\Lambda_{L}^{1} b_{L}+\Lambda_{L}^{2} t_{L}= \Lambda_{L}^{\alpha} q_{L\alpha}\,,
\end{align}
where, abbreviating $\cabb \equiv \cos \theta_{20L}$ and $\sabb \equiv \sin \theta_{20L}$,
\begin{align}
\label{eq:spuri}
\Lambda_{L}^{1} &= \frac{1}{2}\left(\begin{array}{cccccc} 
& & & & \imath e^{i \phi_{20L}}\cabb & \imath\sabb \\
\multicolumn{4}{c}{0_{4 \times 4}} & e^{i \phi_{20L}}\cabb  & \sabb \\
& & & & 0 & 0\\
& & & & 0 & 0\\
\imath e^{i \phi_{20L}}\cabb  &e^{i \phi_{20L}}\cabb  & 0 &0  \\
\imath\sabb & \sabb & 0 & 0 & \multicolumn{2}{c}{0_{2 \times 2}}
\end{array}\right) \,,\nonumber\\
\Lambda_{L}^{2} &=\frac{1}{2}\left(\begin{array}{cccccc} 
& & & & 0 & 0 \\
\multicolumn{4}{c}{0_{4 \times 4}} & 0 & 0\\
& & & & \imath e^{i \phi_{20L}}\cabb  & \imath\sabb \\
& & & &  - e^{i \phi_{20L}}\cabb & -\sabb\\
0& 0& \imath e^{i \phi_{20L}}\cabb & -e^{i \phi_{20L}}\cabb \\
0 & 0 & \imath \sabb & -\sabb & \multicolumn{2}{c}{0_{2 \times 2}}
\end{array}\right) \,,
\end{align}
provide the SO(6) embeddings of the left-handed bottom and top sector, respectively.

As mentioned in the beginning of this section, we assume PC for the coupling of the SM to the composite sector, i.e. we couple the multiplets above to composite fermion resonances $\Psi^{T,t}$ linearly. 
For the following discussion, and in particular for the construction of the pNGB Higgs potential, it is useful to lift the SM fermions to (spurious) SO(6) multiplets by assigning the embedding matrices $\Lambda_L^i$ transformation properties under the full $SO(6)$ global symmetry in an intermediate step, making them so-called spurions of $SO(6)$. This allows to construct the Lagrangian with the help of the spurions from symmetry principles, i.e. an $SO(6)$ symmetry explicitly broken only by setting the spurions back to their actual background values of \eqref{eq:spuri} in the end. Moreover, it is convenient to construct the $SO(6)$-Lagrangian in terms of $SO(5)$ objects as per the CCWZ construction~\cite{Coleman:1969sm,Callan:1969sn,Panico:2015jxa}:
Using the transformation properties of $U$ in \eqref{eq;U.transform}, we can construct the dressed embedding matrices,
\begin{align}
    (\Lambda_L^{\prime \alpha})_{66} = (U^T)_{6I}(U^T)_{6J}(\Lambda_L^\alpha)^{IJ}\,,
    \label{eq:spurion_decomp}\\
    (\Lambda_L^{\prime \alpha})_{a6} = (U^T)_{aI}(U^T)_{6J}(\Lambda_L^\alpha)^{IJ}\,,
    \label{eq:spurion_decomp2}\\
   (\Lambda_L^{\prime \alpha})_{ab} = (U^T)_{aI}(U^T)_{bJ}(\Lambda_L^\alpha)^{IJ}\,,
\end{align}
where $a, b = 1,\dots,5$, obtaining $SO(5)$-singlets, fiveplets and $\mathbf{14}
$-plets\footnote{The bottom-spurion singlet vanishes in this configuration, related to the bottom quark being still massless. To account for a bottom mass, the embedding $Q_L^{20^{\prime}}$ would have to be modified, c.f.~\cite{Panico:2015jxa}, which would however not change our analysis notably.}, and
the resonances can be split similarly.
We then build a formally $SO(5)$ and thus, by virtue of $U$, $SO(6)$ invariant Langrangian, only broken by the spurions aqcuiring their background values. The PC Lagrangian then reads~\cite{Xie:2020bkl}
\begin{align}
    \mathcal{L}_\text{PC} =
    -\!y_L f\! \left(a(\bar{\Psi}_R^{T\prime})^{66} (\Lambda_L^{\prime \alpha})_{66} + b(\bar{\Psi}^{T\prime}_R)^{6a} (\Lambda_L^{\prime \alpha})_{a6} + c (\bar{\Psi}^{T\prime}_R)^{ab}(\Lambda_L^{\prime \alpha})_{ba} \right)q_{L\alpha}  +\! \text{h.c.}\, ,
\end{align} 
where $a,b,c$ are ${\cal O}(1)$ numbers. 

To arrive at the low energy Yukawa Lagrangian $\Lyuk$,
one could integrate out the resonances, yielding form factors, as considered e.g. in ~\cite{Bian:2019kmg,Xie:2020bkl}. 
Alternatively, as explained, one can just construct $SO(5)$ invariants from only the SM fields, spurions and the Goldstone matrix to obtain $\Lyuk$, where the heavy field contributions are absorbed. 
The Yukawa Lagrangian then reads (up to ${\cal O}(1)$ proportionality factors that will drop out in (\ref{eq:Yukf}) below)
\begin{align}
\label{eq:SILH.Yuk}
    \mathcal{L}_\text{Yukawa} =& y_L f\, \bar t_R (\Lambda_L^{\prime \alpha})_{66} q_{L\alpha} + \text{h.c.} \nonumber\\
    =& y_L \, \bar t_R \dfrac{1}{f}\left(-h\sqrt{f^2-h^2-S^2}\sin{\theta_\text{20L}} + \imath h S \cos{\theta_\text{20L}} \right)t_L + \text{h.c.} \nonumber\\
    =& - \sqrt{2}y_L\, \bar q_L  H^c t_R \sin{\theta_\text{20L}}\left(1 - \dfrac{|H|^2}{f^2}-\dfrac{S^2}{2f^2}+\mathcal{O}\!\left(\dfrac{1}{f^4}\right)\right)\nonumber\\ 
    &\quad - \imath \sqrt{2}y_L\cos{\theta_\text{20L}} \dfrac{S}{f} \bar q_L H^c t_R + \text{h.c.}\,.
\end{align}
To make $S$ a real, CP-odd scalar, guaranteeing a $Z_2$-symmetric form of the potential, we chose $\phi_\text{20L}= -\pi/2$ in the second row and in the last row $H^c = \imath\sigma_2 H^\ast = \frac{1}{\sqrt{2}} (h,0)^T$ in unitary gauge.  
We identify $y_t\equiv{\sqrt{2}y_L \sin{\theta_\text{20L}}}$ as the (LO) SM top-Higgs coupling and $\epsilon_Q\equiv\cot{\theta_\text{20L}}$ as the mixing parameter of the embedding, such that, in agreement with \cite{Chala:2017sjk},
\begin{align}
\label{eq:Yukf}
    \mathcal{L}_\text{Yukawa} =
     \dfrac{y_t}{\sqrt{2}} \bar t_R \dfrac{1}{f}\left(-h\sqrt{f^2-h^2-S^2} + \imath \epsilon_Q h S \right)t_L + \text{h.c.}\,.
\end{align}

Summarizing the results so far in the latter parametrization (including the other terms of Eq.~\eqref{eq:Lag}) and generalizing them to three quark families ($i,j=1,..3$) we obtain the effective Lagrangian for the model at hand, up to $D=6$ (see also \cite{Chala:2017sjk}) 
\begin{equation}
\begin{split}
{\cal L}_{\text{(\tiny $\bf 20^\prime\!,\!1$)}}^{\text{\tiny $D\!\leq\!6$}}\supset  & 
(D_\mu H)^\dagger D^\mu H\! +\!\dfrac{1}{2}(\partial_\mu S)^2+\!\dfrac{1}{2f^2}\!\left[\partial_\mu |H|^2+\dfrac{1}{2}\partial_\mu S^2\right]^2 \\
\ 
& -
\sum_{q_R=u_R,d_R}\! (y_q)_{ij}
\bar q^i_L H q^j_R \left(1 + i \epsilon_Q^i \, \frac{S}{f} - \frac{1}{f^2}(|H|^2+S^2/2)\right)+ {\rm h.c.} \\ &\\
& + \frac{S}{16\pi^2f}\left[n_B B_{\mu\nu} \tilde B^{\mu\nu}\!+\!n_W W^{I\mu\nu} \tilde W_{\mu\nu}^I\!+\!n_G G^{a\mu\nu} \tilde G_{\mu\nu}^a\right] - V(H, S)\,,
\end{split}
\end{equation}
where $H\to H^c$ for $q_R\!=\!u_R$ in the second line. 
The first term in the last line corresponds to the WZW Lagragian, with the anomaly coefficients fixed by the coset (with $n_W=n_B,n_G=0$ for $SO(6)/SO(5)$~\cite{Gripaios:2009pe}). These couplings supplement the SM-fermion loops mediating the same transitions, if coupled to $S$. We will now turn to the evaluation of the scalar potential $V(H,S)$.

We can finally construct the potential by using that it is proportional to explicit $SO(6)$ breaking effects, parameterized by the spurions $\Lambda_L^\alpha$. 
Therefore, as mentioned before, we just build formally $SO(6)$ invariant terms consisting of the dressed spurions. Setting them to their background values of (\ref{eq:spuri}) will then lead to the actual scalar potential.
Moreover, to ensure that the terms also respect the SM gauge group, we need an even number of spurions (saturating the SM index).
Assuming a one-scale-one-coupling framework as described, e.g., in \cite{Giudice:2007fh,Panico:2015jxa,Chala:2017sjk}, we can determine the scaling of any potential term by dimensional analysis, 
\begin{align}
    V \propto N_C \dfrac{m_\ast^4}{g_\ast^2}\left(\dfrac{\hbar 
    g_\ast^2}{16 \pi^2}\right)^{\# Loops} \left(\dfrac{y_{L/R}\Lambda}{g_\ast}\right)^{\# spurions}\left(\dfrac{h}{f}\right)^{\# h}\left(\dfrac{S}{f}\right)^{\# S}\,,
\end{align}
where $m_\ast = g_\ast f$ is the resonance mass scale, with $g_\ast$ the coupling of the composite sector, and $N_C$ counts the QCD color multiplicity.

The (one-loop) invariants at LO in $y_L$ read (setting $\hbar=1$) 
\begin{align}
   c_\text{s2}\frac{N_C m_\ast^4}{16 \pi^2}\,\dfrac{y_L^2}{g_\ast^{2}}(\Lambda_L^{\prime \alpha})_{66}(\Lambda_L^{\prime \alpha \dagger})^{66}
    &= c_\text{s2} \frac{N_C g_\ast^2 y_L^2}{16 \pi^2}\,  \left( h^2\left(f^2-h^2-S^2\right)\sabb^2 + h^2 S^2\cabb^2 \right)
    \,,\\
    c_\text{f2}\frac{N_C m_\ast^4}{16 \pi^2}\,\dfrac{y_L^2}{g_\ast^{2}}(\Lambda_L^{\prime \alpha})_{a6}(\Lambda_L^{\prime \alpha \dagger})^{a6}
    &=c_\text{f2}\frac{N_C g_\ast^2 y_L^2}{16 \pi^2}\,\Big(f^4\sabb^2 +\dfrac{f^2}{4}h^2 (\cabb^2-7\sabb^2) + h^4\sabb^2 \nonumber\\
    &\qquad\qquad\qquad - h^2 S^2 (\cabb^2-\sabb^2) + f^2 S^2 (\cabb^2-\sabb^2)\Big)
    \nonumber\,,
    \label{eq:spurions_LO}
\end{align}
where the parameters $c_i \sim {\cal O}(1)$ encode the high energy dynamics and we chose a writing easily comparable with the $\epsilon_Q$ parametrization used in Ref.~\cite{Chala:2017sjk}.\footnote{Note that the 14-plet is not included as, due to $U$ being unitary, it does not induce linearly independent terms.}
This results in the LO potential
\begin{align}
    V\left(h,S\right)= &\frac{N_C g_\ast^2 y_L^2}{16 \pi^2}\,\Big(
    h^2 f^2 \left[ c_\text{s2} \sabb^2 + \dfrac{c_\text{f2}}{4} \left(\cabb^2-7\sabb^2\right)\right]-h^4\sabb^2\left(c_\text{s2}-c_\text{f2}\right) \nonumber\\
    &+h^2 S^2 \left[(c_\text{s2}-c_\text{f2}) \left(\cabb^2-\sabb^2\right)\right]
   + S^2 f^2 c_\text{f2} \left(\cabb^2-\sabb^2\right) \Big)\text{.}    
\end{align}
We can in turn identify the Higgs mass and quartic-coupling, $\mu$ and $\lambda$, which allows us to replace the UV parameters $c_i$ below by connecting them to physical values at low energies.
Note that no quartic $S$ interaction is induced at LO in the spurion expansion. Minimizing with respect to $h$ and $S$ yields the $T=0$ vevs
\begin{align}
    \langle h\rangle \equiv v = \frac{f}{2\sqrt 2} \sqrt{\frac{4  \sabb^2 c_\text{s2}+c_\text{f2} \left(\cabb^2-7\sabb^2\right)}{\sabb^2\left(c_\text{s2}-c_\text{f2}\right)}} =\sqrt{\dfrac{-\mu^2}{\lambda}}, \quad \langle S \rangle = 0
\end{align}
where we already wrote $v$ in terms of $\mu$ and $\lambda$. 
Employing $|H| = h/\sqrt{2}$ (in unitary gauge) finally leads to 
\begin{equation}
\begin{split}
V(H, S) = &\ \mu^2 \left| H\right|^2 + \lambda \left| H\right|^4 + \lambda f^2\left(1-2\frac{v^2}{f^2}\right)\left(\frac{\epsilon_Q^{2}-1}{\epsilon_Q^{2}-3}\right) S^2 \\
 &- \frac12 (\epsilon_Q^{2}-1)\lambda\, S^2 \left| H\right|^2 
 \,.
\end{split}
\end{equation}
Note that, having reverted to the $\epsilon_Q$ parametrization, our result agrees with the one of Ref.~\cite{Chala:2017sjk}. For $\epsilon_Q^2 = 1$, the singlet becomes an exact Goldstone boson, making it massless and not entering the scalar potential.

\subsubsection*{Higher orders in the potential}
To explore if we can obtain a $S^6$ (and $S^4$) term in the model at hand, required for $Z_2$ SNR, in the following we inspect the invariants arising at higher order in spurion insertions (i.e., higher order in $y_L$). 
At next-to-leading order (NLO) and NNLO, in addition to naive powers of LO terms, we need to consider the new structures\footnote{
We denote the coefficients of terms with $n$ powers of the singlet and $m$ powers of the $SO(5)$ fundamental by a subscript sn\hspace{0.3mm}fm. For untilded coefficients, multiplets with adjacent indices are $SU(2)_L$ contracted, while tilded ones correspond to the other possible contraction.} 

\begin{align}
    \tilde c_\text{f4}\frac{N_C m_\ast^4}{16 \pi^2}\ \dfrac{y_L^4}{g_\ast^{4}}&(\Lambda_L^{\prime \alpha})_{a6} (\Lambda_L^{\prime \beta \dagger})^{a6}(\Lambda_L^{\prime \beta})_{b6} (\Lambda_L^{\prime \alpha \dagger})^{b6}
    \,,\nonumber\\
    c_\text{sfsf}\frac{N_C m_\ast^4}{16 \pi^2}\ \dfrac{y_L^4}{g_\ast^{4}}&(\Lambda_L^{\prime \alpha})_{66} (\Lambda_L^{\prime \beta \dagger})^{66}(\Lambda_L^{\prime \beta})_{b6} (\Lambda_L^{\prime \alpha \dagger})^{b6} \,,\nonumber\\
   \tilde c_\text{f6}\frac{N_C m_\ast^4}{16 \pi^2}\ \dfrac{y_L^6}{g_\ast^{6}}&(\Lambda_L^{\prime \alpha})_{a6} (\Lambda_L^{\prime \beta \dagger})^{a6}(\Lambda_L^{\prime \beta})_{b6} (\Lambda_L^{\prime \gamma \dagger})^{b6}(\Lambda_L^{\prime \gamma})_{c6} (\Lambda_L^{\prime \alpha \dagger})^{c6}
    \,,
\end{align} 
as well as corresponding combinations. For simplicity, we present here only the leading contributions for each operator. 
The resulting potential reads
\begin{align}
   \frac{16 \pi^2}{N_C m_\ast^4}\, V\!\left[h,S\right] = &
   \ \tilde\mu_h h^2 
    + \tilde\mu_S S^2 
    + \tilde\lambda_{h4} h^4 
    + \tilde\lambda_{h2S2} h^2 S^2 
    + \tilde\lambda_{S4} S^4 \nonumber
    \\&
    + \tilde\lambda_{h6} h^6     
    + \tilde\lambda_{h4S2} h^4 S^2    
    + \tilde\lambda_{h2S4} h^2 S^4 
    + \tilde\lambda_{S6} S^6    
     +\mathcal{O}((h,S)^8)
\end{align}
\begin{align*}
     \tilde\mu_h &= \dfrac{y_L^{2}}{g_\ast^{2}} \dfrac{1}{f^2} \left[  \sabb^2 c_\text{s2}+  \left(\cabb^2-7\sabb^2\right)\dfrac{c_\text{f2}}{4}\right]
    \\ 
     \tilde\mu_S &=\dfrac{y_L^{2}}{g_\ast^{2}}\dfrac{1}{f^2}  \left(\cabb^2-\sabb^2\right)c_\text{f2}
    \\
    \tilde \lambda_{h4}&=-\dfrac{y_L^{2}}{g_\ast^{2}}\dfrac{1}{f^4}\sabb^2\left(c_\text{s2}-c_\text{f2}\right)\\
     \tilde\lambda_{h2S2}&=\dfrac{y_L^{2}}{g_\ast^{2}}\dfrac{1}{f^4} \left(\cabb^2-\sabb^2\right)(c_\text{s2}-c_\text{f2})
    \\
     \tilde\lambda_{S4} &= \dfrac{y_L^4}{g_\ast^4}\dfrac{1}{2 f^4} \left(\cabb^2-\sabb^2\right)^2\left(2 c_{\text{f4}}+\tilde c_{\text{f4}}\right)
    \\
     \tilde\lambda_{h6} &= \dfrac{y_L^4}{g_\ast^4}\dfrac{1}{4 f^6} \sabb^2\Big[-8\sabb^2 c_{\text{s4}} - \left(\cabb^2-11\sabb^2\right)c_{\text{s2f2}}  \nonumber\\
    &\qquad \qquad \qquad  + 2 \left(\cabb^2-7\sabb^2\right)c_{\text{f4}}+2\left( \cabb^2-5\sabb\right) \tilde c_{\text{f4}}-\left( \cabb^2-9\sabb\right) c_{\text{sfsf}}\Big]
    \\    
     \tilde\lambda_{h4S2} &=
    \dfrac{y_L^4}{g_\ast^4}\dfrac{1}{4 f^6} \left(\cabb ^2-\sabb^2\right) \Big[8\sabb^2 c_{\text{s4}}+ \left(\cabb^2-15\sabb^2\right)c_{\text{s2f2}}\nonumber\\
    &\qquad \qquad \qquad  \qquad -2  \left(\cabb^2-11\sabb^2\right)c_{\text{f4}} -2\left( \cabb^2-7\sabb^2\right) \tilde c_{\text{f4}}+\left( \cabb^2-11\sabb^2\right) c_{\text{sfsf}}\Big]
    \\   
     \tilde\lambda_{h2S4} &= \dfrac{y_L^4}{g_\ast^4}\dfrac{1}{2f^6} \left(\cabb ^2-\sabb^2\right)^2\left(2c_{\text{s2f2}}-4 c_{\text{f4}}-2\tilde c_{\text{f4}}+c_{\text{sfsf}}\right)\\ 
     \tilde\lambda_{S6} &= \dfrac{y_L^6}{g_\ast^6}\dfrac{1}{4f^6} \left(\cabb ^2-\sabb^2\right)^3 \left(4 c_{\text{f6}}+2 \tilde c_{\text{f2f4}}+ \tilde c_{\text{f6}}\right)\,,
\end{align*} 
extending the results of~\cite{Chala:2017sjk}, where we chose again a form that allows for easy translation to the $\epsilon_Q$ parametrization.
We thus find that the $S^4$ and $S^6$ terms are indeed generated at orders $(y_L/g_\ast)^4$ and $(y_L/g_\ast)^6$, respectively.
However, this significant suppression is in tension with a straightforward realization of the SNR scenario and in particular lead to a too large EFT suppression $\Lambda$, see Section~\ref{sec:SNR} and Section~\ref{sec:Match} below. We thus turn to an exploration of the other embeddings mentioned earlier to analyze if the suppression can get lifted. 

%% file: EmbedsTables.tex
\begin{table}[h!]
    \centering
    \caption[$V(h,S)$ terms obtained via spurion analysis for a $SO(6)$ singlet $t_R$ ]{$V(h,S)$ terms obtained via spurion analysis for $t_R$ a $SO(6)$ singlet. Only leading terms are shown and each entry should be multiplied by $N_c m_\ast^4/16\pi^2$, while every $h$ or $S$ comes with a factor $1/f$ and every $y_{L,R}$ with a $1/g_\ast$.
    The coefficients~$c_i$ of the various invariants are just numbered consecutively. Parameters that contribute in the same way in the given approximation were collected into single $c_i$ and we abbreviate $\cabbs=\cth{20L}$, $\sabb^2=\sth{20L}$.\\}
   \label{tab:FermEmbed.1R}
    \small
    \begin{tabular}{|c|c|c|}
    \hline
    \multicolumn{2}{|c|}{}&$\mathbf{ 1}_R$\\ \hline
    \multirow{9}{*}{$\mathbf{ 6}_L$}& $h^2$&\multirow{9}{*}{\color{gray} No S-potential \color{black} }\\
    & $h^4$& \\
    & $h^6$& \\
    & $S^2$& \\
    & $S^4$& \\
    & $S^6$& \\
    & $h^2 S^2$& \\
    & $h^2 S^4$& \\
    & $h^4 S^2$& \\\hline
    \multirow{9}{*}{$\mathbf{ 15_\text{A}}_L$}& $h^2$&\multirow{9}{*}{\color{gray} No top mass \color{black} }\\
    & $h^4$& \\
    & $h^6$& \\
    & $S^2$& \\
    & $S^4$& \\
    & $S^6$& \\
    & $h^2 S^2$& \\
    & $h^2 S^4$& \\
    & $h^4 S^2$& \\\hline
    \multirow{9}{*}{$\mathbf{ 20}^\prime_L$}&$h^2$&$y_L^2 \left(\sabb^2 c_1+\frac{1}{4}\left(4\cabbs-3\right) c_2\right)$
    \\\cline{2-3}
    & $h^4$& $-y_L^2 \sabb^2\left( c_1- c_2\right)$\\\cline{2-3}
    & $h^6$& $\frac{1}{4} y_L^4 \sabb^2\left(- c_4+2 c_5+2 c_6- c_7-2\sabb^2\left(4 c_3-6 c_4+8 c_5+6 c_6-5 c_7\right)\right)$\\\cline{2-3}
    & $S^2$& $y_L^2  \cabbs c_2$
    \\\cline{2-3}
    & $S^4$& $\frac{1}{2}y_L^4 \cabbs^2\left(2 c_5+ c_6\right)$\\\cline{2-3}
    & $S^6$& 
    $\frac{1}{4} y_L^6 \cabbs^3 4 c_8$
    \\\cline{2-3}
    & $h^2 S^2$& $y_L^2 \cabbs\left( c_1- c_2\right)$
    \\\cline{2-3}
    & $h^2 S^4$& $-\frac{1}{2} y_L^4 \cabbs^2\left(-2 c_4+4 c_5+2 c_6- c_7\right)$
    \\\cline{2-3}
    & $h^4 S^2$& $\frac{1}{4}y_L^4 \cabbs\left(4\sabb^2 \left(2 c_3-4 c_4+6 c_5+4 c_6-3 c_7\right)-\left(- c_4+2 c_5+2 c_6- c_7\right)\right)$ 
    \\\hline
    \end{tabular}
    \normalsize
\end{table}
\begin{table}[h!]
    \centering
   \caption[$V(h,S)$ terms obtained via spurion analysis for $t_R$ in a $SO(6)$ sextuplet]{
   $V(h,S)$ terms obtained via our spurion analysis for $t_R$ in a $SO(6)$ sextuplet. Only leading terms are shown and each entry should be multiplied by $N_c m_\ast^4/16\pi^2$, while every $h$ or $S$ comes with a factor $1/f$ and every $y_{L,R}$ with a $1/g_\ast$. The $\mathbf{ 15}_L$ model does not generate a CPV-inducing term and is thus displayed in gray. Parameters that contribute in the same way in the given approximation were collected into single $c_i$ and we abbreviate $\cabbs=\cth{6R}$ and $\sabb^2=\sth{6R}$, while the other angle is set to zero, respectively. As discussed in the text, going to finite values of the displayed angle changes the hierarchy of terms for the $\mathbf{20^\prime}_L$ model.\\}
\label{tab:FermEmbed.6R}
    \small
    \begin{tabular}{|c|c|c|}
    \hline
    \multicolumn{2}{|c|}{}&$\mathbf{ 6}_R$\\ \hline
    \multirow{9}{*}{$\mathbf{ 6}_L$}& $h^2$&$\frac{1}{2} y_L^2 c_1 - y_R^2 \sabb^2 c_2 $\\\cline{2-3}
    & $h^4$& $\frac{1}{4} y_L^4c_4 -\frac{1}{2} y_L^2 y_R^2  \sabb^2 \left(-2 c_3\right) + y_R^4 \sabb^4 c_5$\\\cline{2-3}
    & $h^6$& $\frac{1}{8}\left(y_L^6 c_{8}-2  y_L^4 y_R^2 \sabb^2 c_6+4 y_L^2 y_R^4 \sabb^4 c_7-8 y_R^6 \sabb^6  c_{9} \right)$\\\cline{2-3}
    & $S^2$& $ y_R^2\cabbs c_2$\\\cline{2-3}
    & $S^4$& $ y_R^4\cabbs^2 c_5$\\\cline{2-3}
    & $S^6$& $ y_R^6\cabbs^3 c_{9}$\\\cline{2-3}
    & $h^2 S^2$& $\frac{1}{2} y_R^2 \cabbs \left(-2 y_L^2 c_3 -4 y_R^2 \sabb^2 c_5 \right)$\\\cline{2-3}
    & $h^2 S^4$& $\frac{1}{2} y_R^4 \cabbs^2 \left(y_L^2 c_7 - 6  y_R^2 \sabb^2 c_{9}\right)$\\\cline{2-3}
    & $h^4 S^2$& $\frac{1}{4} y_R^2 \cabbs \left(y_L^4 c_6-4  y_L^2 y_R^2 \sabb^2 c_7 +12 y_R^4 \sabb^4 c_{9}\right)$
    \\\hline
    \multirow{9}{*}{$\mathbf{ 15_\text{A}}_L$}& $h^2$&
    \color{gray}$\frac{1}{4}y_L^2  c_1 - y_R^2 \sabb^2   c_2 $\\\cline{2-3}
     & $h^4$&
    \color{gray} $\frac{1}{16}y_L^4\left(c_4+c_5\right)-\frac{1}{4}y_L^2 y_R^2\sabb^2 c_3+y_R^4 \sabb^4 c_6$\\\cline{2-3}
    & $h^6$&\color{gray} \tiny $\frac{1}{64} h^6 \left(16 c_9 y_L^2 y_R^4 \sabb^4-4 \left(c_7+c_8\right) y_L^4 y_R^2 \sabb^2+\left(c_{10}+c_{11}-8 c_{12}\right) y_L^6-64 c_{13} y_R^6 \sabb^6\right)$\\\cline{2-3}
    & $S^2$&\color{gray} $y_L^2  c_1 + y_R^2  \cabbs  c_2 $\\\cline{2-3}
    & $S^4$&\color{gray} $ y_L^2 y_R^2 \cabbs c_3+\frac{1}{2}y_L^4\left(2c_4+c_5\right) + y_R^4 \cabbs^2 c_6$\\\cline{2-3}
    & $S^6$&\color{gray} 
    $\frac{1}{2} \left( y_L^4 y_R^2 \cabbs \left(2 c_7+c_8\right)+ y_L^2 y_R^4 (\mathfrak{c}_{4\theta}+1) c_9+ y_L^6\left(2 c_{10}+c_{11}\right)+2  y_R^6 \cabbs^3 c_{13}\right)$
    \\\cline{2-3}
    & $h^2 S^2$&\color{gray} $\frac{1}{4} y_L^4\left(2c_4+c_5\right)+\frac{1}{4} y_L^2 y_R^2 (3\cabbs-2)  c_3-2 y_R^4 \sabb^2\cabbs  c_6$\\\cline{2-3}
    & $h^2 S^4$&\color{gray}\tiny 
    $\frac{1}{8}  \left(2  y_L^4 y_R^2 (2 \cabbs-1)\left(2 c_7+c_8\right)+2y_L^2 y_R^4 \cabbs (5 \cabbs-4) c_9 +3 y_L^6 \left(2 c_{10}+c_{11}\right)-24 y_R^6 \sabb^2 \cabbs^2 c_{13} \right)$
    \\\cline{2-3}
    & $h^4 S^2$&\color{gray}\tiny 
    $\frac{1}{16} \left(y_L^4 y_R^2 \left(c_7 (5 \cabbs-4)c_7+(3 \cabbs-2)c_8 \right)+8  y_L^2 y_R^4 \sabb^2 (1-2 \cabbs)c_9+ y_L^6\left(3 c_{10}+2 c_{11}\right)+48  y_R^6 \sabb^4 \cabbs c_{13}\right)$
    \\\hline
    \multirow{9}{*}{$\mathbf{ 20_\text{A}^\prime}_L$}
    & $h^2$     & $y_L^2 \frac{1}{4} c_2- y_R^2 \sabb^2 c_3 $\\\cline{2-3}
    & $h^4$     & $\frac{1}{64} \left(4y_L^4 \left( c_9+ c_{10}\right)+2y_L^2y_R^2\left(\left(3\cabbs-1\right)c_5-4\sabb^2 c_6\right)+ y_R^4\left(3\cabbs-1\right)^2 c_{12}\right)$\\\cline{2-3}
    & $h^6$     & $\frac{1}{64} \left(y_L^6 c_{16} - 4y_L^4y_R^2\sabb^2 c_{13} +16 y_L^2y_R^4\sabb^4 c_{14} -64y_R^6\sabb^6 c_{15}\right)$\\\cline{2-3}
    & $S^2$     & $y_L^2 c_2 + y_R^2 \cabbs c_3 $ \\\cline{2-3}
    & $S^4$     & $\frac{1}{2}y_L^4 \cabbs^2 \left(2 c_9+c_{10}\right)$\\\cline{2-3}
    & $S^6$     & $\frac{1}{4} y_L^6 4 c_{17}$\\\cline{2-3}
    & $h^2 S^2$ & $ y_L^2 \left( c_1- c_2\right)$\\\cline{2-3}
    & $h^2 S^4$ & $\frac{1}{2} y_L^4\left(2 c_8-4 c_9-2 c_{10}+c_{11}\right)$\\\cline{2-3}
    & $h^4 S^2$ & $\frac{1}{4} y_L^4 \left(c_8-2 c_9-2 c_{10}+c_{11}\right)+ \frac{1}{8} y_L^2 y_R^2 \left(3\cabbs-1\right)\left(c_4-c_5\right)$ \\\hline
    \end{tabular}
    \normalsize
\end{table}
\begin{table}[h!]
    \centering
    \caption[$V(h,S)$ terms obtained via spurion analysis for $t_R$ in a $\mathbf{ 15}$ of $SO(6)$]{$V(h,S)$ terms obtained via our spurion analysis for $t_R$ in a $\mathbf{ 15}$ of $SO(6)$.
    Only leading terms are shown and each entry should be multiplied by $N_c m_\ast^4/16\pi^2$, while every $h$ or $S$ comes with a factor $1/f$ and every $y_{L,R}$ with a $1/g_\ast$. Parameters that contribute in the same way in the given approximation were collected into single $c_i$ and we abbreviate $\cabbs=\cth{15R}$ and $\sabb^2=\sth{15R}$,  while the other angle is set to zero, respectively. As discussed in the text, going to finite values of the displayed angle changes the hierarchy of terms for the $\mathbf{20^\prime}_L$ model.\\}
\label{tab:FermEmbed.15R}
    \begin{tabular}{|c|c|c|}
    \hline
    \multicolumn{2}{|c|}{}&$\mathbf{ 15}_R$\\ \hline
    \multirow{9}{*}{$\mathbf{ 6}_L$}& $h^2$&\multirow{9}{*}{\color{gray} No S-potential \color{black} }\\
    & $h^4$& \\
    & $h^6$& \\
    & $S^2$& \\
    & $S^4$& \\
    & $S^6$& \\
    & $h^2 S^2$& \\
    & $h^2 S^4$& \\
    & $h^4 S^2$& \\\hline
    \multirow{9}{*}{$\mathbf{ 15_\text{A}}_L$}&
    $h^2$&$\frac{1}{4}y_L^2  c_1+\frac{1}{8} y_R^2\left(3\cabbs-1\right) c_2$
    \\\cline{2-3}
    & $h^4$& $\frac{1}{64} \left(2 y_L^2 y_R^2 \left((3 \cabbs-1)c_3 -4  \sabb^2c_4\right)+4y_L^4 \left(c_5+c_6\right) + y_R^4 (1-3 \cabbs)^2 c_7\right)$
    \\\cline{2-3}
    & $h^6$& \tiny $\frac{1}{512} \bigg(4 y_L^4 y_R^2 \left(\left(3 c_8+3 c_9+2 c_{12}+2 c_{13}\right) \cabbs-c_8-c_9-2 c_{12}-2 c_{13}\right)$\\
    & & \quad\tiny$+2 y_L^2 y_R^4 (3 \cabbs-1) \left(\left(3 c_{10}+2 c_{11}\right) \cabbs-c_{10}-2 c_{11}\right)+8 \left(c_{14}+c_{15}+c_{16}\right) y_L^6+c_{17} y_R^6 (3 \cabbs-1)^3\bigg)$
    \\\cline{2-3}
    & $S^2$& $y_L^2  c_1$
    \\\cline{2-3}
    & $S^4$& $y_L^4 \frac{1}{2}\left(2c_5+c_6\right)$
    \\\cline{2-3}
    & $S^6$& $\frac{1}{4}y_L^6 \left(4 c_{14}+2 c_{15}+c_{16}\right)$
    \\\cline{2-3}
    & $h^2 S^2$& $\frac{1}{32}  y_L^2 \left(8 \left(2 c_5+c_6\right) y_L^2+\left(4 c_3+c_4\right) y_R^2 (3 \cabbs-1)\right)$
    \\\cline{2-3}
    & $h^2 S^4$& \tiny $\frac{1}{64} y_L^4 \left(12 \left(4 c_{14}+2 c_{15}+c_{16}\right) y_L^2+\left(8 c_8+4 c_9+2 c_{12}+c_{13}\right) y_R^2 (3 \cabbs-1)\right)
    $\\\cline{2-3}
    & $h^4 S^2$& 
    \tiny $\frac{1}{256}y_L^2 \bigg(y_L^2 y_R^2 \left(\left(48 c_8+24 c_9+22 c_{12}+14 c_{13}\right)\cabbs-16 c_8-8 c_9-18 c_{12}-10 c_{13}\right)$\\
    & &\tiny\quad$+8 \left(6 c_{14}+4 c_{15}+3 c_{16}\right) y_L^4+\left(4 c_{10}+c_{11}\right) y_R^4 (1-3\cabbs)^2\bigg)$
    \\\hline
    \multirow{9}{*}{$\mathbf{ 20_\text{A}^\prime}_L$}
    & $h^2$     & $\frac{1}{4} y_L^2  c_2 +\frac{1}{8} y_R^2 \left(3\cabbs - 1\right) c_3$\\\cline{2-3}
    & $h^4$     & $\frac{1}{64}\left(4 y_L^4 \left( c_9+ c_{10}\right) + 2 y_L^2 y_R^2\left(\left(3\cabbs-1\right)c_5-4\sabb^2 c_6\right)+y_R^4\left(3\cabbs-1\right)^2 c_{12}\right)$\\\cline{2-3}
    & $h^6$     & \tiny $\frac{1}{256}\left(y_L^6 4 c_{17} + y_L^4 y_R^2 \left(2\left(3\cabbs-1\right)c_{13} + 8\sabb^2 c_{14}\right)\right.$\\
    &           & \tiny $\left.\qquad+y_L^4 y_R^2 \left(\left(3\cabbs-1\right)^2 c_{15} + 4\sabb^2 \left(3\cabbs-1\right) c_{16}\right)+\frac{1}{2} y_R^6 \left(3\cabbs-1\right)^2 c_{18}\right)$\\\cline{2-3}
    & $S^2$     & $y_L^2 c_2 $ \\\cline{2-3}
    & $S^4$     & $y_L^4 \frac{1}{2}\left(2c_{9}+c_{10}\right)$\\\cline{2-3}
    & $S^6$     & $\frac{1}{4} y_L^6 4 c_{19}$\\\cline{2-3}
    & $h^2 S^2$ & $ y_L^2 \left( c_1- c_2\right)$\\\cline{2-3}
    & $h^2 S^4$ & $\frac{1}{2} y_L^4 \left(2 c_8-4 c_9-2 c_{10}+c_{11}\right)$\\\cline{2-3}
    & $h^4 S^2$ & $\frac{1}{4} y_L^4 \left(c_8-2 c_9-2 c_{10}+c_{11}\right)+\frac{1}{4}y_L^2 y_R^2\left(3\cabbs-1\right)\left(c_4-c_5\right)$\\\hline
    \end{tabular}
\end{table}
\begin{table}[h!]
\caption[$V(h,S)$ terms obtained via spurion analysis for $t_R$ in a $\mathbf{20^\prime_\text{AB}}$ of $SO(6)$]{$V(h,S)$ terms obtained via our spurion analysis for $t_R$ in a $\mathbf{20^\prime_\text{AB}}$ of $SO(6)$ $(\theta_{20R2}=0)$. 
This configuration is chosen in order to reproduce a heavy top quark and simultaneously a CPV-inducing term, combining viably with the $\mathbf{20^\prime_\text{A}}_L$ such as to generate a $S^6$ term at NLO and fitting our SNR criteria.  
Only leading terms are shown and each entry should be multiplied by $N_c m_\ast^4/16\pi^2$, while every $h$ or $S$ comes with a factor $1/f$ and every $y_{L,R}$ with a $1/g_\ast$. Parameters that contribute in the same way in the given approximation were collected into single $c_i$ and we abbreviate $\cabbs=\cth{20R1}$ and $\sabb^2=\sth{20R1}$, while $\theta_{15L}, \theta_{20L}=0$.
The smaller left-handed embeddings do not fit the SNR coupling pattern in this choice of angles -- for a discussion of nonzero $\theta_{20R2}$, which does allow SNR also in those cases, see main text.\\}
\label{tab:FermEmbed.20R}
\centering
\footnotesize
\begin{tabular}{|c|c|c|}
    \hline
    \multicolumn{2}{|c|}{}&$\mathbf{ 20^\prime}_R$\\ \hline
    \multirow{9}{*}{$\mathbf{ 6}_L$}& 
    $h^2$& $\frac{1}{2}y_L^2 c_1+\frac{1}{40}y_R^2  \left(11\cabbs-9\right)c_3$
    \\\cline{2-3}
    & $h^4$&  $\frac{1}{20}y_R^2 \cabb^2\left( c_2- c_3\right)$
    \\\cline{2-3}
    & $h^6$&  $\frac{1}{800}\cabb^2\left(20y_L^2 y_R^2 \left(-2 c_4\right)+y_R^4\left(11\cabbs-9\right)\left( c_6-2 c_7\right)\right)$
    \\\cline{2-3}
    & $S^2$& $y_R^2 \left(2\sabb^2 c_2+\frac{1}{5}\left(7\cabbs-3\right) c_3\right)$
    \\\cline{2-3}
    & $S^4$& $y_R^2 \frac{1}{5}\left(7\cabbs-3\right)\left( c_2- c_3\right)$
    \\\cline{2-3}
    & $S^6$&  $\frac{4}{25}y_R^4\left(7\cabbs-3\right)\left( c_6-2 c_7+\sabb^2\left(5 c_5-6 c_6+7 c_7\right)\right)$
    \\\cline{2-3}
    & $h^2 S^2$& $y_R^2 \frac{2}{5}\left(2\cabbs-3\right)\left( c_2- c_3\right)$
    \\\cline{2-3}
    & $h^2 S^4$& \tiny$ \frac{1}{10} y_L^2 y_R^2 \left(7 \cabbs-3\right)\left(-2 c_4\right)$
    \\& &\tiny$+\frac{1}{200}y_R^4\bigg(-64  \sabb^2\left(25  c_5-29  c_6+33  c_7\right) +\mathfrak{s}_{2\theta}^2\left(320  c_5-461  c_6+602  c_7\right)-56 \left( c_6-2  c_7\right)\bigg)$ 
    \\\cline{2-3}
    & $h^4 S^2$& \tiny$ \frac{1}{5}  y_L^2y_R^2 \left(2 \cabbs-3\right)\left(-2 c_4\right)+\frac{1}{200} y_R^4 \left(2 \sabb^2\left(5  c_5-28  c_6+51  c_7\right) -2  \left(49 \cabbs-51\right)\left( c_6-2  c_7\right)\right)$ \\\hline
    \multirow{9}{*}{$\mathbf{ 15_\text{A}}_L$}
    & $h^2$&$\frac{1}{4}y_L^2  c_1+\frac{1}{40}y_R^2\left(11\cabbs-9\right) c_3$
    \\\cline{2-3}
    & $h^4$& $\frac{1}{20}y_R^2\cabb^2\left( c_2- c_3\right)$
    \\\cline{2-3}
    & $h^6$& $\frac{1}{800}y_R^2 \cabb^2 \left(10 y_L^2 c_4+ y_R^2 \left(11 \cabbs-9\right) \left(c_6-2 c_7\right) \right) $
    \\\cline{2-3}
    & $S^2$& $y_L^2  c_1+y_R^2 \left(2\sabb^2 c_2+\frac{1}{5}\left(7\cabbs-3\right) c_3\right)$\\\cline{2-3}
    & $S^4$& $\frac{1}{5}y_R^2\left(7\cabbs-3\right)\left( c_2- c_3\right)$\\\cline{2-3}
    & $S^6$& $\frac{1}{25}y_R^2 \left(7 \cabbs-3\right) \left(5 c_4 y_L^2+y_R^2 \left(2 \left(\cabbs-1\right)\left(-5 c_5+6 c_6-7 c_7\right) +4 \left(c_6-2 c_7\right)\right)\right)$\\\cline{2-3}
    & $h^2 S^2$& $\frac{2}{5}y_R^2\left(2\cabbs-3\right)\left( c_2- c_3\right)$\\\cline{2-3}
    & $h^2 S^4$& 
    \tiny$\frac{1}{200} y_R^2 \bigg(10 y_L^2 \left(23 \cabbs-27\right) c_4 $\\
    & &
    \tiny
    $+y_R^2 \sabb^2 \left(-64 \left(25 c_5-29 c_6+33 c_7\right) +\mathfrak{s}_{2\theta}^2\left(320 c_5-461 c_6+602 c_7\right) -56 \left(c_6-2 c_7\right)\right)\bigg)
    $\\\cline{2-3}
    & $h^4 S^2$& \tiny$\frac{1}{200}  y_R^2 \left(5 y_L^2 \left(9 \cabbs-11\right) c_4-y_R^2 \left(2 \left(49 \cabbs-51\right) \left(c_6-2 c_7\right) -2 \mathfrak{s}_{2\theta}^2 \left(5 c_5-28 c_6+51 c_7\right) \right)\right)$
    \\\hline
    \multirow{9}{*}{$\mathbf{ 20^\prime_\text{A}}_L$}
    &$h^2$&$y_L^2 \frac{1}{4} c_2+\frac{1}{40}y_R^2\left(11\cabbs-9\right) c_4$\\\cline{2-3}
    & $h^4$& $\frac{1}{20}y_R^2\cabb^2\left( c_3- c_4\right)$\\\cline{2-3}
    & $h^6$& $\cabb^2\left(\frac{1}{80}y_L^2 y_R^2 \left( c_7- c_8\right)+\frac{1}{800}y_R^4\left(11\cabbs-9\right)\left( c_{17}-2 c_{18}\right)\right)$
    \\\cline{2-3}
    & $S^2$& $y_L^2  c_2+\frac{1}{5}y_R^2\left(5 c_3-3 c_4+\cabbs\left(-5 c_3+7 c_4\right)\right)$\\\cline{2-3}
    & $S^4$& $\frac{1}{5}y_R^2 \left(7\cabbs-3\right)\left( c_3- c_4\right)$\\\cline{2-3}
    & $S^6$& $\frac{1}{5}y_L^2y_R^2\left(7\cabbs-3\right)\left( c_7- c_8\right)$\\
    & &$-\frac{2}{25}y_R^4\left(7\cabbs-3\right)\left(-5 c_{16}+4 c_{17}-3 c_{18}+\cabbs\left(5 c_{16}-6 c_{17}+7 c_{18}\right)\right)$\\\cline{2-3}
    & $h^2 S^2$& $y_L^2 \left( c_1- c_2\right)+\frac{2}{5}y_R^2\left(2\cabbs-3\right)\left( c_3- c_4\right)$\\\cline{2-3}
    & $h^2 S^4$& \tiny$y_L^4 \frac{1}{2}\left(2 c_{12}-4 c_{13}-2 c_{14}+ c_{15}\right)$\\
    & &
    \tiny$-\frac{1}{20}y_L^2 y_R^2 \left(-20  c_5+12  c_6+47  c_7-39  c_8+44  c_9-24  c_{10}+\cabbs\left(20  c_5-28  c_6-43  c_7+51  c_8-56  c_9+36  c_{10}\right)\right)$\\
    & &
    \tiny$+\frac{1}{400}y_R^4\left(64 \cabbs\left(25  c_{16}-29  c_{17}+33  c_{18}\right)  +\mathfrak{c}_{4\theta}\left(-320  c_{16}+461  c_{17}-602  c_{18}\right) -1280  c_{16}+1283  c_{17}-1286  c_{18}\right)$\\\cline{2-3}
    & $h^4 S^2$&\tiny $\frac{1}{4}y_L^4\left( c_{12}-2  c_{13}-2  c_{14}+ c_{15}\right) $\\
    & &
    \tiny$+ \frac{1}{40} y_L^2 y_R^2 \left(\cabbs\left(11  c_6+9  c_7-20  c_8+17  c_9-17  c_{10}\right) -9  c_6-11  c_7+20  c_8-23  c_9+23  c_{10}\right)$\\
    & &
    \tiny$+\frac{1}{200}y_R^4 \left(-98 \cabbs\left( c_{17}-2  c_{18}\right) +\mathfrak{c}_{4\theta}\left(-5  c_{16}+28  c_{17}-51  c_{18}\right) +5  c_{16}+74  c_{17}-153  c_{18}\right)$\\\hline
    \end{tabular}
\end{table}

%% file: Appendix.tex
\section{Yukawa Terms of Different Embeddings}\label{appx:ferm.allYukawa}
We provide here the explicit results for the Yukawa couplings in the various models considered. For simplicity, we abbreviate the dressed fermion multiplets as $\left(Q_L^{\prime {m}}\right)_{ij} = \left(\Lambda_L^{\prime {m} \alpha}\right)_{ij} q_{L\alpha}$, and similar for the right-handed fermions. We obtain
\begin{align}
    \Lyuk^{( 6,  6)} &= \dfrac{y_L y_R^\ast}{g_\ast}f \left(\bar Q_L^{\prime {6}}\right)_6 \left(t_R^{\prime {6}}\right)_6 M_1 + \text{h.c.}\\
    &= -\dfrac{y_L y_R^\ast}{\sqrt{2}g_\ast f}\bar{t}_L\left[ h \sqrt{f^2-h^2-S^2} \sin \left(\theta _{6 R}\right)+i h S \cos \left(\theta _{6 R}\right)\right]M_1 t_R + \text{h.c.} \nonumber
    \\
    \Lyuk^{( 15,  6)} &= \dfrac{y_L y_R^\ast}{g_\ast}f \left(\bar Q_L^{\prime {15}}\right)_{6a} \left(t_R^{\prime {6}}\right)_a M_5 + \text{h.c.} \\
    &= -\dfrac{y_L y_R^\ast}{2 g_\ast f} \bar{t}_L h \cos \left(\theta _{6 R}\right)  M_{5} t_R + \text{h.c.}  \nonumber
    \\
    \Lyuk^{( 20^\prime,  6)} &= \dfrac{y_L y_R^\ast}{g_\ast}f \left[\left(\bar Q_L^{\prime {20^\prime}}\right)_{66} \left(t_R^{\prime {6}}\right)_6 M_1 + \left(\bar Q_L^{\prime {20^\prime}}\right)_{6a} \left(t_R^{\prime {6}}\right)_a M_5 \right]+ \text{h.c.}\\   
    & = \dfrac{y_L y_R^\ast}{2g_\ast f^2}  \bar t_L \left[- f^2 h  \cos\left(\theta_{6R}\right) M_5 - 2 h S^2   \cos\left(\theta_{6R}\right)\left(M_1-M_5\right)\right. \nonumber \\ & \left. \qquad\qquad\quad +2 i  h S  \sqrt{f^2-h^2-S^2} \sin\left(\theta_{6R}\right) \left(M_1-M_5\right) \right] t_R + \text{h.c.}
\end{align}
\begin{align}
    \Lyuk^{( 15,  15)} &= \dfrac{y_L y_R^\ast}{g_\ast} f \left(\bar Q_L^{\prime {15}}\right)_{6a} \left(t_R^{\prime {15}}\right)_{a6} M_5+ \text{h.c.}
    \\
    &= \dfrac{y_L y_R^\ast}{4g_\ast f} \bar t_L   \left[\sqrt{2} h\sqrt{f^2-h^2-S^2}\sin (\theta_{15R}) +i h S \cos (\theta_{15R})\right]M_5  t_R+ \text{h.c.}\nonumber\\
    \Lyuk^{( 20^\prime,  15)} &= \dfrac{y_L y_R^\ast}{g_\ast} f \left(\bar Q_L^{\prime {20^\prime}}\right)_{6a} \left(t_R^{\prime {15}}\right)_{a6} M_5+ \text{h.c.}
    \\
    &= \dfrac{y_L y_R^\ast}{4 g_\ast f}\bar t_L \left[h \sqrt{f^2-h^2-S^2} \sin\left(\theta_{15R}\right)+i  h S \cos\left(\theta_{15R}\right)\right] M_5  t_R + \text{h.c.}\nonumber
\end{align}
\begin{align}
    \Lyuk^{( 6,  20^\prime)} 
    &= \dfrac{y_L y_R^\ast}{g_\ast} f \left[\left(\bar Q_L^{\prime {6}}\right)_{6} \left(t_R^{\prime {20^\prime}}\right)_{66} M_1+\left(\bar Q_L^{\prime {6}}\right)_{a} \left(t_R^{\prime {20^\prime}}\right)_{a6} M_5\right]+ \text{h.c.}
    \\
    &= \dfrac{y_L y_R^\ast}{g_\ast f^2}\bar t_L \bigg[
    -f^2 h  \Big\{
    \left(\dfrac{1}{2\sqrt{10}}e^{i \phi _{20R1}}\cos \left(\theta _{20R1}\right)+\sqrt{\dfrac{3}{5}}\sin \left(\theta _{20R1}\right)\sin \left(\theta _{20R2}\right)\right) M_5
    \nonumber\\
    &\qquad\qquad\qquad\qquad-\dfrac{\sqrt{15}}{6} \sin \left(\theta _{20R1}\right)\sin \left(\theta _{20R2}\right) M_1\Big\}
    \nonumber\\
    &\qquad\qquad\quad+ h  S \sqrt{f^2-h^2-S^2} e^{i \phi _{20R2}}\cos \left(\theta _{20R2}\right)\sin \left(\theta _{20R1}\right)\left(M_5-M_1\right)
    \nonumber\\
    &\qquad\qquad\quad  +\frac{h S^2 \left(\sqrt{2} e^{i \phi _{20R1}}\cos \left(\theta _{20R1}\right)-\sqrt{3} \sin \left(\theta _{20R1}\right) \sin \left(\theta _{20R2}\right)\right)}{\sqrt{5}}\left(M_5-M_1\right)
    \nonumber\\
    &\qquad\qquad\quad
    -\frac{h^3 \left(\sqrt{2} e^{i \phi _{20R1}}\cos \left(\theta _{20R1}\right)+4 \sqrt{3} \sin \left(\theta _{20R1}\right) \sin \left(\theta _{20R2}\right)\right)}{4 \sqrt{5}}\left(M_5-M_1\right)\bigg]t_R\nonumber\\
    &\qquad\qquad
    + \text{h.c.}
    \nonumber\\
    &\overset{here}{=}
    -\dfrac{y_L y_R^\ast}{g_\ast f^2}\bar t_L \bigg[ f^2 h \dfrac{\cos\left(\theta _{20R1}\right)}{2\sqrt{10}}M_5+ i\cdot h S \sqrt{f^2-h^2-S^2} \sin \left(\theta _{20R1}\right)\left(M_5-M_1\right)
     \nonumber\\
    &\qquad\qquad\quad+  \left(\sqrt{\dfrac{2}{5}}h S^3 -\dfrac{1}{2\sqrt{10}}h^3\right) \cos \left(\theta _{20R1}\right)\left( M_5- M_1\right)\nonumber\bigg]
    + \text{h.c.}
    \nonumber
    \\
    \Lyuk^{( 15,  20^\prime)} &= \dfrac{y_L y_R^\ast}{g_\ast} f \left(\bar Q_L^{\prime {15^\prime}}\right)_{6a} \left(t_R^{\prime {20^\prime}}\right)_{a6} M_5+ \text{h.c.}\nonumber\\
    &= i \dfrac{y_L y_R^\ast}{4 g_\ast f} \bar t_L e^{-i \phi _{15 \text{L}}}\bigg[\sqrt{2} h e^{i \phi _{20R2}}\sin \left(\theta_{20R1}\right)\cos \left(\theta_{20R2}\right) \sqrt{f^2-h^2-S^2}\\
    &\qquad\qquad\quad+ \sqrt{5} h S e^{i \phi _{20R1}}\cos \left(\theta_{20R1}\right)\bigg]M_5 t_R+ \text{h.c.} \nonumber\\
    &\overset{here}{=}
      -\dfrac{y_L y_R^\ast}{4 g_\ast f}\bar t_L\left[\sqrt{2} \sin \left(\theta _{20R1}\right)\cdot h \sqrt{f^2-h^2-S^2} + i\sqrt{5}\cdot h S\cos  \left(\theta _{20R1}\right)\right]t_R M_5 + \text{h.c.}\nonumber
    \\
    \Lyuk^{( 20^\prime,  20^\prime)} &= \dfrac{y_L y_R^\ast}{g_\ast} f \left[\left(\bar Q_L^{\prime {20^\prime}}\right)_{66} \left(t_R^{\prime {20^\prime}}\right)_{66} M_1+\left(\bar Q_L^{\prime {20^\prime}}\right)_{6a} \left(t_R^{\prime {20^\prime}}\right)_{a6} M_5\right]+ \text{h.c.} \label{eq:Yuk20}\\
    & =\dfrac{y_L y_R^\ast}{g_\ast f^3} \bar t_L \bigg[-f^2 h\sqrt{f^2-h^2-S^2}\frac{\sin \left(\theta_{20R1}\right)  M_5 }{2 \sqrt{2}}-i f^2 h S\frac{3  \cos \left(\theta_{20R1}\right) M_5}{4 \sqrt{5}}
    \nonumber\\
    &\qquad\qquad\quad - h S^2  \sqrt{f^2-h^2-S^2} \sqrt{2}  \sin \left(\theta_{20R1}\right)\left(M_1-M_5\right) 
    \nonumber\\
    &\qquad\qquad\qquad + \dfrac{i}{2\sqrt{5}}\left( h^3 S -4 h S^3 \right)\cos \left(\theta_{20R1}\right)\left(M_1-M_5\right) 
    \bigg] t_R +\text{h.c.}\, ,\nonumber
\end{align}
where $M_1,M_5$ are the form factors associated to invariants from $SO(5)$ singlets and fiveplets. These expressions allow to explicitly see the generation of the top mass as well as to estimate the size of CPv induced in terms of the mixing angles.
\newpage